\documentclass[pra,twocolumn,amsmath,amssymb,floatfix,reprint]{revtex4-1}

\usepackage{graphicx}
\usepackage{dcolumn}
\usepackage{bm}
\usepackage[usenames,dvipsnames]{xcolor}
\usepackage{mciteplus}
\usepackage{braket}

\begin{document}

\title{Quantum engines with interacting Bose-Einstein condensates}
\author{Julián Amette Estrada$^{1,2}$\email{julianamette@df.uba.ar},
  Franco Mayo$^{1,3}$\email{Loremipsum@physique.ens.fr},
  Augusto J. Roncaglia$^{1,3}$\email{Loremipsum@physique.ens.fr} \&
  Pablo D. Mininni$^{1,2}$\email{mininni@df.uba.ar}}
  
\affiliation{$^1$Universidad de Buenos Aires, Facultad de Ciencias Exactas y Naturales, Departamento de Física, Ciudad Universitaria, 1428 Buenos Aires, Argentina,}
\affiliation{$^2$CONICET - Universidad de Buenos Aires, Instituto de F\'{\i}sica Interdisciplinaria y Aplicada (INFINA), Ciudad Universitaria, 1428 Buenos Aires, Argentina.}
\affiliation{$^3$CONICET - Universidad de Buenos Aires, Instituto de F\'{\i}sica de Buenos Aires (IFIBA), Ciudad Universitaria, 1428 Buenos Aires, Argentina.}

\date{\today}

\begin{abstract}
We consider a quantum Otto cycle with an interacting Bose-Einstein condensate at finite temperature. We present a procedure to evolve this system in time in three spatial dimensions, in which closed (adiabatic) strokes are described by the Gross-Pitaevskii equation, and open (isochoric) strokes are modeled using a stochastic Ginzburg-Landau equation. We analyze the effect on the thermodynamic efficiency of the strength of interactions, the frequency of the harmonic trap, and the temperatures of the reservoirs. The efficiency has little sensitivity to changes in the temperatures, but decreases as interactions increase. However, stronger interactions allow for faster cycles and for substantial increases in power.
\end{abstract}
\maketitle

\section{Introduction}
\label{sec:introduction}
Quantum thermodynamics~\cite{vinjanampathy2016quantum,binder2018thermodynamics} has emerged as a captivating field of research that bridges the fundamental principles of quantum mechanics and the laws of thermodynamics. In recent years, there has been a growing interest in the study of quantum thermal machines~\cite{myers2022quantum,mitchison2019quantum,bhattacharjee2021quantum, fialko2012isolated}, which are devices that utilize quantum systems to convert heat into work and vice versa. In this context, how genuine quantum effects, such as quantum coherence~\cite{camati2019coherence, dann2020quantum}, correlations \cite{hewgill2018quantum}, and measurements~\cite{elouard2017extracting,elouard2018efficient,jordan2020quantum} can be exploited to improve the performance of these machines has been the subject of intense study. In addition to these theoretical studies, several experimental implementations of different quantum thermodynamic cycles have also been implemented using single quantum systems, such as trapped ions and atoms~\cite{rossnagel2016single,abah2012single,von2019spin}. 

Quantum many-body systems have also been proposed as the working medium for engines and refrigerators. 
In this context, 
Bose-Einstein condensates (BECs) have emerged as a prominent candidate due to their remarkable macroscopically observable quantum properties and controllability. BECs, are formed by cooling a gas of bosonic particles to extremely low temperatures and
are characterized by a high degree of coherence, where a significant fraction of the particles occupy the same quantum state. The precise control  achieved over BECs through techniques such as laser cooling and magnetic trapping, allows for the manipulation of their properties and opens up exciting possibilities for exploring quantum thermodynamics.
Recent works have designed various engines that utilize BECs to extract work. For instance, in~\cite{myers2022boosting} it was considered an endoreversible Otto cycle with  non-interacting Bose gas, showing that the power output can be enhanced in a regime when the working medium is in the BEC phase. In~\cite{reyes2023carnot}, an interacting BEC engine was explored, and their performance was addressed through the experimental determination of the equation of state. Interacting BECs were also considered in~\cite{keller2020feshbach, li2018efficient}, for a cycle working with a particle reservoir at zero temperature, and where the interaction strength between atoms is controlled by Feshbach resonances. In~\cite{niedenzu2019quantized}, a strategy for using  a mixture of two atomic gases as a quantum refrigerator is outlined. Additionally, in~\cite{gluza2021quantum} there is a proposal on building quantum engines using one-dimensional ultracold gases and illustrates its use in the cooling process.  

In this paper, we study a quantum Otto cycle using a three dimensional interacting Bose-Einstein condensate as a working medium. To do so we perform direct numerical simulations in which the closed (adiabatic) strokes are described by the Gross-Pitaevskii equation (GPE) and the open (isochoric) strokes, that occur at finite temperature, are modeled using a stochastic Ginzburg-Landau equation. This approach allows us to obtain the complete dynamics of the BEC during the whole cycle, and provides a highly detailed description of the quantum-many body engine. Therefore, we not only obtain the whole thermodynamic description of the system, but we also are able to track the evolution of the different contributions to the energy, as well as the state of the BEC consistently. 
We aim to uncover the underlying principles governing the efficiency, power output, and other relevant features of these machines. By employing advanced theoretical models and numerical simulations, we systematically analyze how various parameters, such as the interaction strength, frequency of the harmonic trap, and reservoirs temperatures, affect the performance of a quantum Otto cycle.

\section{Methods} 
\label{sec:methods}

\subsection{Thermodynamic cycle}

We will consider a finite-time Otto cycle using an interacting Bose-Einstein condensate as a working medium. The thermodynamic cycle starts with the gas at temperature $T_h$, in an spherical trap with frequency $\omega_h$. The first stroke is an adiabatic expansion that turns $\omega_h$ to $\omega_c$, with $\omega_h > \omega_c$, thus expanding the condensate. During the second stroke the system is put in contact with an external cold source, and the gas cools down to reach a thermalized state at temperature $T_c < T_h$ in an isochoric process. The third stroke is an adiabatic compression, changing the trap potential from $\omega_c$ to $\omega_h$. Finally, in the last stroke the system undergoes an isochoric process in contact with a hot source at temperature $T_h$. Thus, the cycle is completely described by prescribing the time taken during each of the strokes, respectively $\tau_e$ (for the expansion), $\tau_\textrm{cold}$, $\tau_c$ (for the compression), and $\tau_\textrm{hot}$; together with the time dependence of the trap potential  during the adiabatic strokes. In the following, we will use the same expansion and contraction times so that $\tau_{e,c} = \tau_e = \tau_c$. 

We can define $W_c$ and $W_e$ respectively as the works extracted in the compression and the expansion, and $Q_h$ as the heat absorbed in the isochoric hot process,
\begin{align}
    W_{c,e} &= E_{c,e}^{(i)} - E_{c,e}^{(f)}, \\
    Q_h &= E_e^{(f)} - E_c^{(i)}.
\end{align}
Here $E$ is the system total energy, the subindices $e$ and $c$ denote respectively the expansion and contraction strokes, and the superindices $i$ and $f$ denote respectively the initial and final states of these strokes. The efficiency of a heat engine is defined as the net yielded work ($W = W_c + W_e$)  divided by the absorbed heat. 
In practice, extended systems display heat fluctuations, and variations in the work as the cycle is repeated. The efficiency of the cycle is defined as 
\begin{equation}
    \eta = \frac{W}{Q_h}.
\end{equation}
This efficiency will fluctuate in different realizations of the cycle, so one usually is concerned with the mean efficiency.  
For non-interacting condensates in the adiabatic regime, the efficiency reduces to the Otto efficiency~\cite{myers2022boosting}
\begin{equation}
    \eta_O = 1 - \frac{\omega_c}{\omega_h}.
    \label{eq:etaO}
\end{equation}

\subsection{Adiabatic evolution}
We will describe the state of the Bose-Einstein condensate in terms of a single wave function $\psi(\mathbf{r},t)$. For the expansion and the contraction strokes we solve numerically the Gross-Pitaevskii equation (GPE) with a time-dependent harmonic trapping potential $V({\bf r},t)$,
\begin{equation}
     i \hbar \frac{\partial \psi (\mathbf{r},t)}{\partial t} = \left[  -\frac{\hbar^2 \boldsymbol{\nabla}^2}{2 m} + g |\psi (\mathbf{r},t)|^2 + V(\mathbf{r},t)\right]\psi (\mathbf{r},t).
   \label{eq:GPE}
\end{equation}
Here $m$ is the atomic mass, the interaction is controlled by $g = 4 \pi a \hbar^2/m$, and $a$ is the s-wave scattering length. The spherical potential is  given by $V(\mathbf{r},t) = m \omega^2(t) (x^2+y^2+z^2)/2$, and the frequency $\omega(t)$ during the adiabatic strokes changes linearly in time from the initial to the final value.
In order to evaluate the total energy of the condensate we will consider the Hamiltonian associated to Eq. \eqref{eq:GPE}:
\begin{multline}
\mathcal{H}[\psi,\psi^*] = \int d^3r' \left[ \frac{\hbar^2}{2m} |\boldsymbol{\nabla} \psi|^2 + \frac{g}{2} |\psi|^4 + V(\mathbf{r},t) |\psi|^2 \right],
\label{eq:Hamiltonian}
\end{multline}
where the star denotes the complex conjugate.

\subsection{Thermal baths}

During the isochoric strokes the system is coupled to a thermal bath, and in principle it can exchange both particles and energy. Under these conditions the possible equilibria will be characterized by a volume $\mathcal{V}$, a chemical potential $\mu$, and a temperature $T$. The probability of these equilibrium states is then given by the Grand canonical ensemble,
\begin{equation}
    \mathbb{P} = \frac{e^{-\beta (\mathcal{H} - \mu \mathcal{N})}}{\mathcal{Z}},
    \label{eq:GC}
\end{equation}
where $\beta=1/(k_B T)$, $k_B$ is the Boltzmann constant, $\mathcal{Z}$ is the Grand canonical partition function, and $\mathcal{N}$ is the number of particles in the system.

The evolution of the system towards these equilibria, while in contact with a thermal bath at temperature $T$, can be done in terms of the approach described in \cite{Krstulovic2011, AmetteEstrada2022b}. Thus, by adding white-noise to Eq.~\eqref{eq:GPE} we solve the following stochastic Ginzburg-Landau equation:
\begin{eqnarray}
    \frac{\partial \psi}{\partial t} &=& \left[  \frac{\hbar}{2 m}\boldsymbol{\nabla}^2 - \frac{g}{\hbar} |\psi |^2 - V(\mathbf{r}) + \frac{\mu}{\hbar} \right]\psi + \nonumber \\ 
    {} &&
    \sqrt{\frac{2}{\mathcal{V} \hbar \beta}} \zeta (\mathbf{r},t) .
    \label{eq:GLET}
\end{eqnarray}
Where $\zeta (\mathbf{r},t)$ is a delta correlated random process such that $\left< \zeta (\mathbf{r},t) \zeta^* (\mathbf{r}',t')\right> = \delta (\mathbf{r} - \mathbf{r}') \delta (t-t')$, and the factor $\sqrt{2 / (\mathcal{V} \hbar \beta)}$ controls the amplitude of the fluctuations through the temperature $T$. This equation can be obtained by performing a Wick rotation $t \rightarrow it$ to Eq.~\eqref{eq:GPE}, and by adding both the chemical potential and the delta correlated random forcing term. In the absence of forcing this equation evolves into solutions that are stationary solutions of GPE \cite{Nore1997}. Note that the Ginzburg-Landau equation is also used to study non-isolated dissipative dynamics, e.g., in superconductivity \cite{Severino_2022}.


We can explicitly verify that the solutions of Eq.~\eqref{eq:GLET} result in equilibria compatible with Eq.~\eqref{eq:GC}. Defining the free energy $F = \mathcal{H} - \mu \mathcal{N}$, Eq.~\eqref{eq:GLET} can be written as a Langevin equation for the evolution of each Fourier mode of $\psi$ \cite{Krstulovic2011},
\begin{equation}
    \frac{\partial \hat{\psi}(\bm{k},t)}{\partial t} = - \frac{1}{\mathcal{V} \hbar} \frac{\partial F }{\partial \hat{\psi}^* (\bm{k},t)} + \sqrt{\frac{2}{\mathcal{V} \hbar \beta}} \hat{\zeta}(\bm{k}, t) ,
    \label{eq:langevin_fourier}
\end{equation}
where $F = F[\{\hat{\psi} (\bm{k},t), \hat{\psi}^* (\bm{k},t)\}]$ (i.e., it is a functional of the set of Fourier amplitudes of $\psi$, where a Galerkin truncation up to a maximum wave-number is applied to the set of Fourier modes such that $|\bm{k}| < k_\textrm{max}$). The resulting stochastic process has a total state probability $\mathbb{P}[\{\hat{\psi} (\bm{k},t), \hat{\psi}^* (\bm{k},t)\} ]$ whose evolution is described by a corresponding multivariate Fokker-Planck equation \cite{VanKampen}
\begin{equation}
    \frac{\partial \mathbb{P}}{\partial t} = \sum_{\bm{|k| < k_{\textrm{max}}}} \frac{\partial}{\partial \hat{\psi}_{\bm{k}}} \left[ \frac{1}{\mathcal{V} \hbar} \frac{\partial F}{\partial \hat{\psi}_{\bm{k}}^*} \mathbb{P} + \frac{1}{\mathcal{V} \hbar \beta} \frac{\partial \mathbb{P}}{\partial \hat{\psi}_{\bm{k}}^*} \right] + \ {\rm c.c.} 
\end{equation}
where $\hat{\psi}_{\bm k}$ is shorthand for $\hat{\psi} (\bm{k},t)$, and c.c. denotes the complex conjugate. This equation evolves into the Grand canonical distribution in Eq.~\eqref{eq:GC} provided that $\beta F$ is positive defined. Thus, by integrating numerically Eq.~\eqref{eq:GLET} we can evolve the system towards states with different temperatures $T$ under the Grand canonical constraints.

For the isochoric strokes, the system evolves at constant volume $\mathcal{V}$ and fixed number of particles $\mathcal{N}$ (or equivalently, at fixed mean density $\Bar{\rho}$ in the total volume that contains the gas). This corresponds to working on the Canonical ensemble, and can be achieved by solving Eq.~\eqref{eq:GLET} coupled with \cite{Krstulovic2011,AmetteEstrada2022b}
\begin{equation}
    \frac{\partial \mu}{\partial t} = - \gamma (\Bar{\rho} - \rho_m) .
\end{equation} 
This equation adjust the chemical potential such that the mean density $\Bar{\rho}$ remains close to the target mean density in the trap $\rho_m
$; $\gamma$ is a parameter that controls how fast convergence to the desired mean density takes place.

It is worth noting that there exist other formulations to describe condensates at finite temperature, such as the stochastic Gross-Pitaevskii equation \cite{Gardiner_2002, Calzetta_2007}, or coupled kinetic equations \cite{Zaremba_1999}. While the method used here generates the correct thermal states and has long been used to study the dynamics of dissipative systems at finite temperature \cite{Schmid_1966}, a stochastic Gross-Pitaevskii or kinetic formulation could better describe nonequilibrium dynamics although at a larger computational cost. For a comparison between these methods see \cite{Berloff_2014}.

\subsection{Energetics}

The total energy of the system can be decomposed into several components that provide information on excited  ordered and disordered modes in the gas, such as potential and internal energies, or a compressible kinetic energy that can be associated to sound waves and phonons. To this end we use the Madelung transformation,
\begin{equation}
    \psi (\mathbf{r},t) = \sqrt{\rho(\mathbf{r},t)/m} \, e^{i S(\mathbf{r},t)},
    \label{eq:Madelung}
\end{equation}
which maps GPE to the Euler equation for an isentropic, compressible and irrotational gas with an extra term that accounts for quantum pressure \cite{Nore1997}. This allows for a continuum medium description of the system. In Eq.~\eqref{eq:Madelung} the transformation $\rho(\mathbf{r},t)$ is the fluid mass density, and $S(\mathbf{r},t)$ is the phase of the order parameter. Using the momentum density
\begin{equation}
\bm{j}(\mathbf{r},t) = - \frac{i \hbar}{2} \left( \psi^* \boldsymbol{\nabla} \psi - \psi \boldsymbol{\nabla} \psi^* \right),
\end{equation}
the gas velocity can then be defined as $\mathbf{v} (\mathbf{r},t) =  \bm{j}(\mathbf{r},t)/\rho(\mathbf{r},t)=(\hbar/m) \boldsymbol{\nabla} S (\mathbf{r},t)$.

Thus, in terms of the fluid mass density, the total energy of the system per unit volume (see Eq.~\eqref{eq:Hamiltonian}) can be decomposed as 
\begin{equation}
E = E_{\rm k} + E_{\rm q} + E_{\rm int} + E_{\rm V},
\end{equation}
where the kinetic energy is $E_{\rm k} = \langle \rho v^2 \rangle/2$, the quantum energy is $E_{\rm q}= \hbar^2/(2m^2) \langle (\nabla \sqrt{\rho})^2 \rangle$, the gas internal (or interaction) energy is $E_{\rm i}= g/(2m^2) \langle \rho^2 \rangle$, and the trap potential energy is $E_{\rm V} = \langle \rho V \rangle$. In all cases the angle brackets denote volume average. Using the Helmholtz decomposition $(\sqrt \rho {\bf v})=(\sqrt \rho {\bf v})^{\rm (c)}+ (\sqrt \rho {\bf v})^{\rm (i)}$ \cite{Nore1997}, where the superindices c and i denote respectively the compressible and incompressible components (i.e., such that $\nabla \cdot (\sqrt \rho {\bf v})^{\rm (i)}=0$), the kinetic energy can be further decomposed into the compressible $E_{\rm k}^{\rm (c)}$ and incompressible $E_{\rm k}^{\rm (i)}$ kinetic energy components. This decomposition is used to study classical compressible gasses \cite{KidaOrszag1990}, as well as quantum fluids \cite{Shukla2019, AmetteEstrada2022, AmetteEstrada2022b}, and thus provides information that can be compared with the classical picture of thermal engines. 

\subsection{Numerical methods \label{sec:numerics}}

We solve numerically Eqs.~\eqref{eq:GPE} and \eqref{eq:GLET} to simulate respectively the adiabatic and isochoric strokes of the cycle. 
In order to do so we use a pseudospectral Fourier-based method in a spatial grid of $N^3 = 64^3$ points, with the $2/3$ rule for dealiasing, a fourth-order Runge-Kutta method for the time evolution of GPE, and an Euler time stepping method for the stochastic Ginzburg-Landau equation. In all cases we use the parallel code GHOST, which is publicly available \cite{Mininni2011}, in a cubic domain of dimensions $[-\pi,\pi]L \times [-\pi,\pi]L \times [-\pi,\pi]L$, so that the domain has length $2\pi L$  (with $L$ a unit length). To deal with the non-periodic trapping potential in the Fourier representation, while avoiding Gibbs phenomenon, we use a continuation method as described in \cite{Fontana_2020, AmetteEstrada2022}.

Results are shown in units of a characteristic speed $U$ (the speed of sound), the unit lenght $L$ (proportional to the condensate mean radius), and a unit mean density $\rho_0$. Temperatures are written in units of $T_\lambda$, the condensate critical temperature (see the Appendix for its estimation, and for the range of temperatures considered in this study). Except when explicitly stated (e.g., when we study the effect of varying $T_h$), we consider $T_h \approx 0.012 T_\lambda$ and $T_c \approx 0.003 T_\lambda$. Thus, the simulations have $T \ll T_\lambda$. The speed of sound is $c = (g \rho_0/m)^{1/2} = 1U$ and the condensate healing length is $\xi = \hbar/(2m \rho_0 g)^{1/2} = 0.0707 L$, except in simulations in which we artificially decrease the interaction strength. In most simulations we use trapping frequencies $\omega_c \approx 0.334638 \, U/L$ and $\omega_h = 0.337613 \, U/L$. These frequencies are chosen close enough to reduce the computational cost of performing the slow expansions and contractions, and we indicate explicitly when other values of $\omega_c$ and $\omega_h$ are used. Quantities can be scaled using dimensional values for $U$, $L$, and $M$. In experiments typical dimensional values are $L\approx 10^{-4}$ m and $c  = U \approx 2 \times 10^{-3}$ m/s \cite{White2014}. This results in $\xi \approx 1.12 \times 10^{-6}$ m and a mean trap frequency $\omega \approx 4.7$ Hz. For the typical mass of a gas of $^{87}$Rb atoms in a BEC, peak densities of $\approx 10^{13}$ cm$^{-3}$ atoms are also compatible with our simulations and with experiments \cite{Henn2008}.

For a given set or parameters, each cycle is repeated 4 times. This results in several values for the energies $E_i^{(j)}$ (with $i=c$ or $e$, and $j=i$ or $f$), and thus for $W$ and $Q_h$ at the end of each cycle. To compute efficiencies we assume these quantities have a Gaussian distribution, and use a Montecarlo method to generate a random set of $W$ and $Q_h$ values compatible with the fluctuations observed in the 4 explicitly integrated cycles. From these values, the distribution of the efficiency $\eta$ and its mean value are finally obtained.

Each realization of the cycle is performed with the following protocol: Given a state at temperature $T_h$ (which can be generated for the first cycle by integrating the stochastic Ginzburg-Landau equation, or can be the result of the final state of a previous cycle), we integrate the expansion stroke using GPE. The frequency of the trap is decreased linearly in time from $\omega_h$ to $\omega_c$ with a time step $dt = 2.5 \times 10^{-3} L/U$; the length of this simulation depends on the speed of the expansion. When the expansion finishes, the system is evolved towards the lower temperature $T_c$ using the stochastic Ginzburg-Landau equation. Time integration is performed until the system reaches a stationary regime. Then, the contraction is integrated using GPE with a linear ramp in the trap frequency from $\omega_c$ to $\omega_h$, with the same $dt$ and total time as in the expansion. Finally, the system is again coupled to the hot source at temperature $T_h$, and integrated using the stochastic Ginzburg-Landau equation until a new stationary state is reached.

\section{Results}

\subsection{Analysis of the system evolution on a cycle}

We first analyze the evolution of the system in each stroke, and the evolution of the different energy components, for the set of parameters introduced in Sec.~\ref{sec:numerics}. Then, we evaluate the efficiency of the cycle in terms of the variation of these parameters.

\begin{figure}[tb]
    \centering
    \includegraphics[width=\columnwidth]{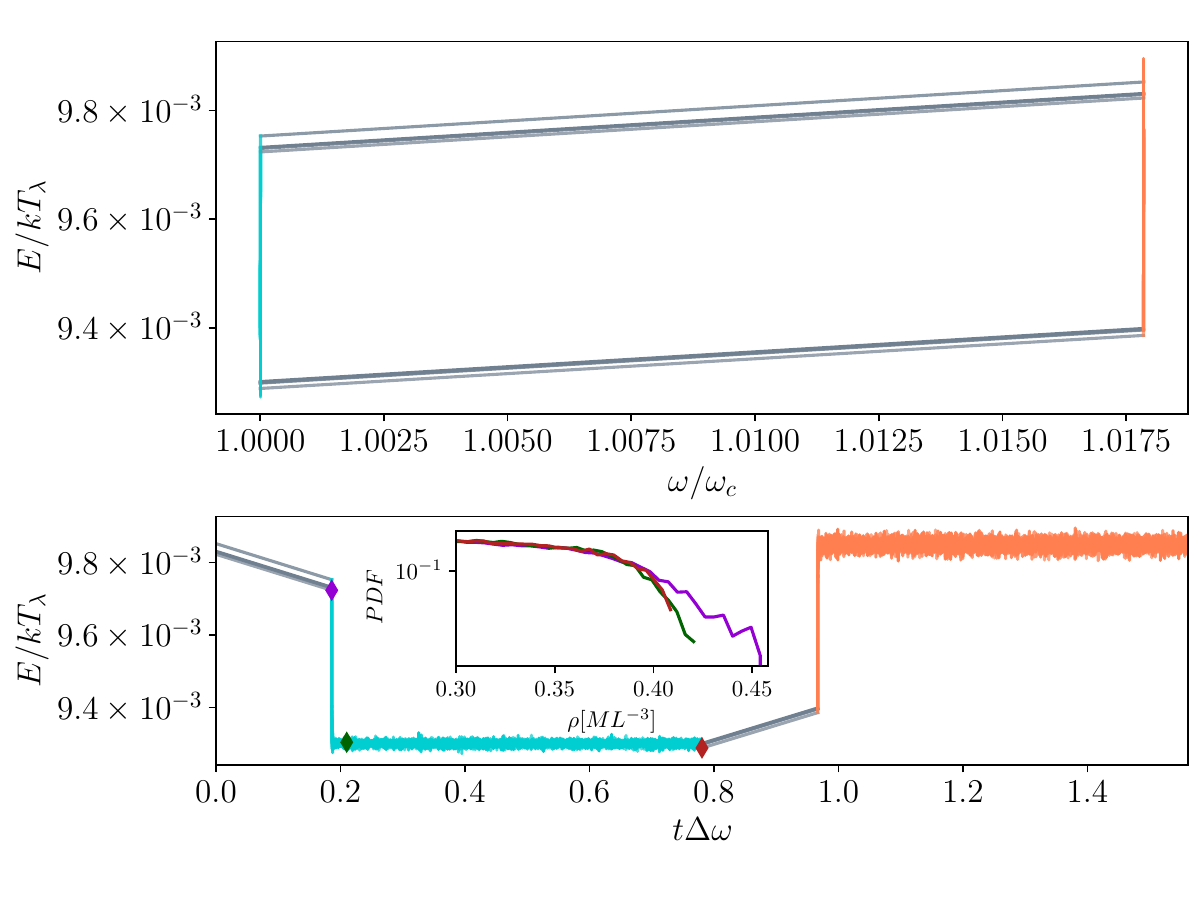}
    \caption{Top panel: Energy as a function of the trap frequency, $\omega$, for several consecutive cycles. Expansions and contractions are plotted in gray, cooling and heating strokes are in cyan and orange, respectively. Bottom panel: Time evolution (with time in units of the inverse of $\Delta \omega = \omega_h - \omega_c$) of the total energy in the same cycles, setting time $t=0$ for all at the beginning of the expansion. The inset shows the probability density function (PDF) of the condensate mass density in the trap at different times; colors of the lines match the times of the diamond markers in the main figure.}
    \label{fig:cycle plot}
\end{figure}

A diagram of several consecutive cycles in the energy-frequency plane, and their time evolution (with the time $t=0$ set arbitrarily at the beginning of each cycle expansion, and with the time given in units of the inverse of the frequency difference $\Delta \omega = \omega_h - \omega_c$), are shown in Fig.~\ref{fig:cycle plot}. The numerical simulations agree with the usual picture of an Otto cycle. An abrupt change in energy can be seen as soon as the condensate gets in contact with the thermal sources. The full dynamics allows us to see fluctuations, both in the energy after the adiabatic phases, as well as those produced by the thermalization process. These fluctuations also result in slightly different values of the energy along each of the expansions and contractions in the different cycles.

During the isochoric strokes a long integration time is necessary for the system to thermalize at the new temperature. The inset in Fig.~\ref{fig:cycle plot} shows the tails of the probability density functions (PDFs) of the mass density in the trap at different times after the system is coupled to the cold source. At early times, as the condensate is still hot,   the PDF displays strong tails, associated to strong fluctuations in the mass density. Shortly after these regions with strong fluctuations disappear and as the condensate cools down, the PDFs converge to new stationary solutions with weaker tails. Similar results are obtained for the evolution with the hot source. In the next subsection we vary the strength of the interaction in the BEC and verify that even for weak interactions the system thermalizes. 
We also ensured that the isochoric branches were integrated long enough to achieve stationary and accurate convergence of the PDFs.

\begin{figure}[tb]
    \centering
    \includegraphics[width=\columnwidth]{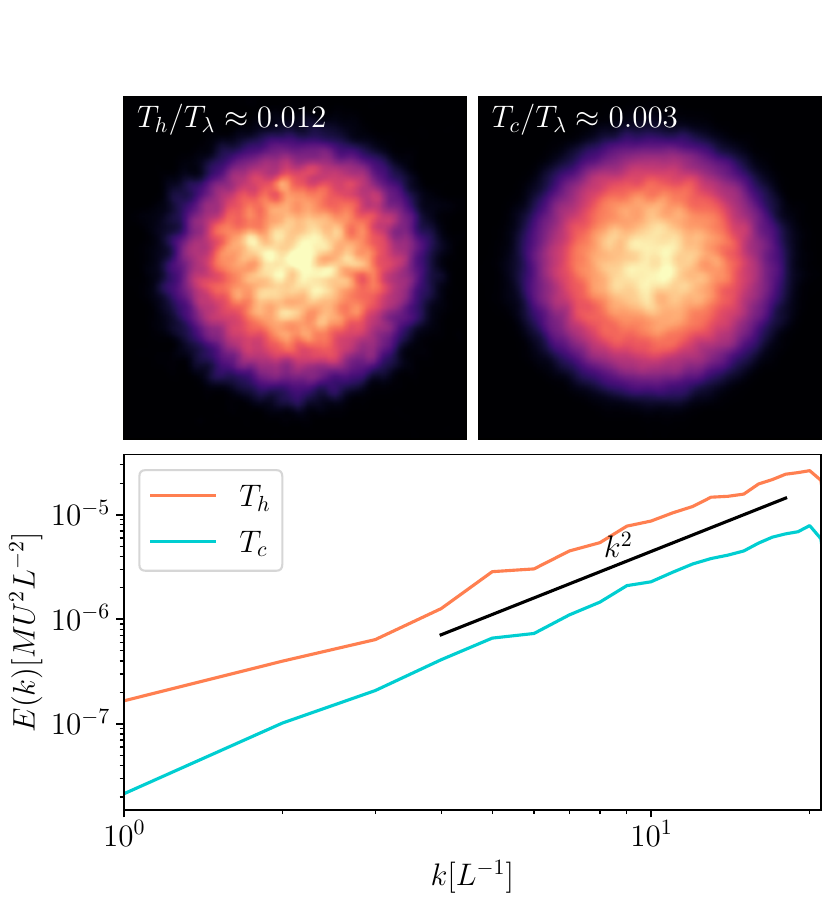}
    \caption{Top panel: Two-dimensional slices of the mass density in the $xy$ plane, $\rho(x,y,z=0)$, for temperatures $T_h/T_{\lambda} \approx 0.012$ and $T_c/T_{\lambda} \approx 0.003$. Bottom panel: Spectrum of the compressible kinetic energy in the hot and cold cases. A $\sim k^2$ power law corresponding to equipartition of compressible three-dimensional modes is shown as a reference.}
    \label{fig:spectrums}
\end{figure}

The final states of the isochoric strokes are shown in Fig.~\ref{fig:spectrums}. The top panels show the mass density in a two-dimensional slice in the $xy$ plane, at the end of the hot and cold branches solved with the stochastic Ginzburg-Landau equation. At higher temperature the gas displays stronger fluctuations in the mass density at the center of the trap as well as in the borders of the condensate where irregularities can be seen. Density fluctuations are associated with more energy in compressible modes (sound waves or phonons), and with an increase in the quantum energy (caused by gradients in the mass density). The bottom panel shows the compressible kinetic energy spectrum in both cases. Note that this spectrum measures the energy in sound waves. 
Two interesting features are worth mentioning. First, the spectra are proportional to a $k^2$ power law, which corresponds to the equipartition of energy in 3D modes (i.e., thermalisation). Second, the amplitude of sound waves increases with temperature.

\begin{figure}[tb]
    \centering
    \includegraphics[width=\columnwidth]{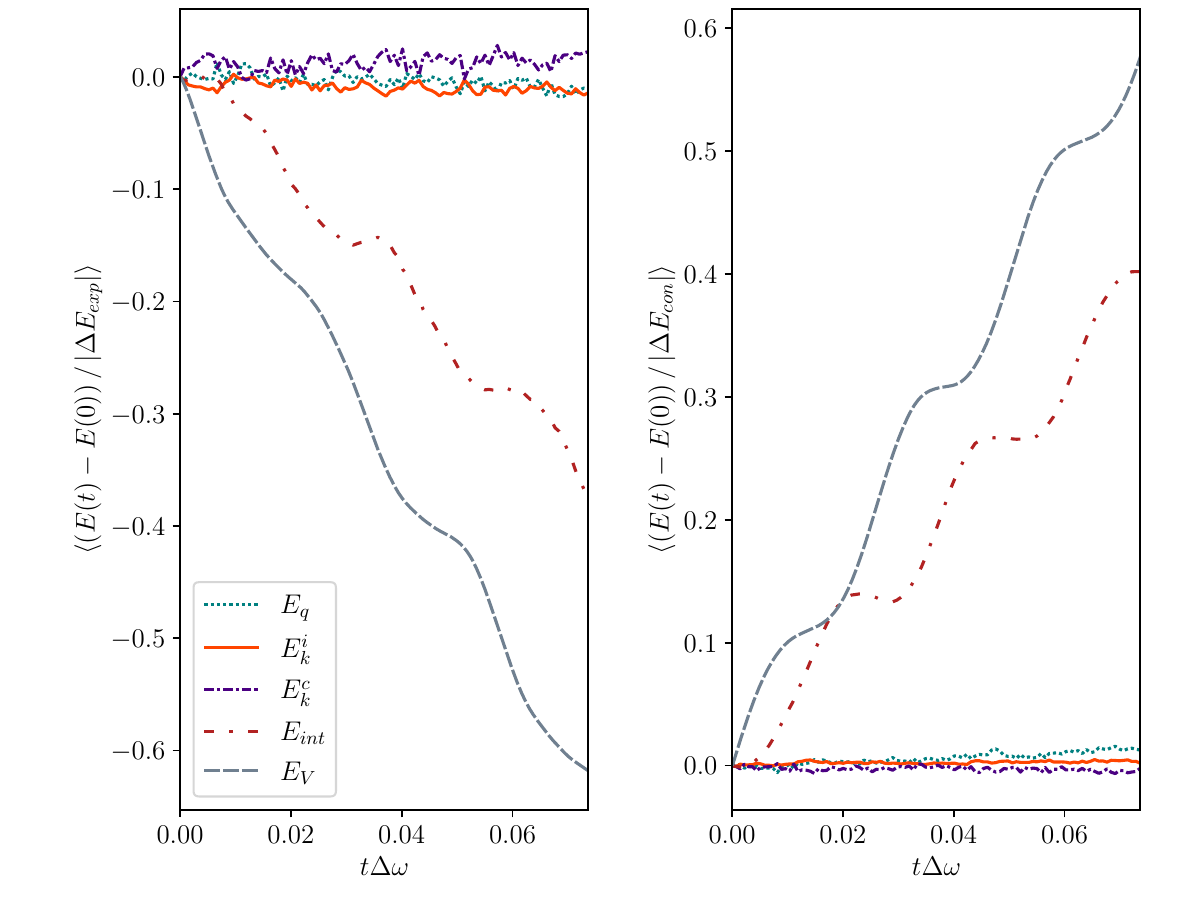}
    \caption{Time evolution of the different energy components, averaged over an ensemble of four cycles, during the adiabatic expansion (left) and contraction (right). The value of each energy component at the beginning of the strokes (here arbitrarily labeled as $t=0$) is subtracted from the energies, and the energy variations are then normalized by the absolute value of the total energy difference during the entire stroke.}
    \label{fig:energy components evolution}
\end{figure}

Now, we will analyse the behaviour of the energy components during the adiabatic strokes. Naturally, this depends on whether we consider a compression or an expansion, as well as on the speed of the stroke. For the sake of clarity we now consider shorter strokes than in Fig.~\ref{fig:cycle plot} (i.e., faster compressions and expansions), as they result in more evident effects. Figure \ref{fig:energy components evolution} shows the time evolution of the different energy components, averaged over four cycles. The energy variations are normalized by the absolute value of the total energy difference during the stroke. During the expansion, the interaction and trap potential energies decay rapidly as the condensate expands. Both quantities also oscillate with the frequency of the breathing mode of the condensate in the trap and have, due to the nature of each energy, almost opposite phases. Meanwhile, the compressible kinetic energy grows as sound waves are excited during the expansion. The incompressible and quantum energies remain almost constant. During the compression, the interaction and trap potential energies grow as the condensate contracts. In this case the compressible kinetic energy remains almost constant, with a small increase of the quantum energy as density gradients grow due to the contraction.

\subsection{Efficiency}

Now, we will analyze the efficiency of the cycle. In particular, we will be focused on its behaviour in terms of the speed of the expansion and compression (i.e. $\tau_{e,c}$), the temperature, and the interaction strength.

\begin{figure}[tb]
    \centering
    \includegraphics[width=\columnwidth]{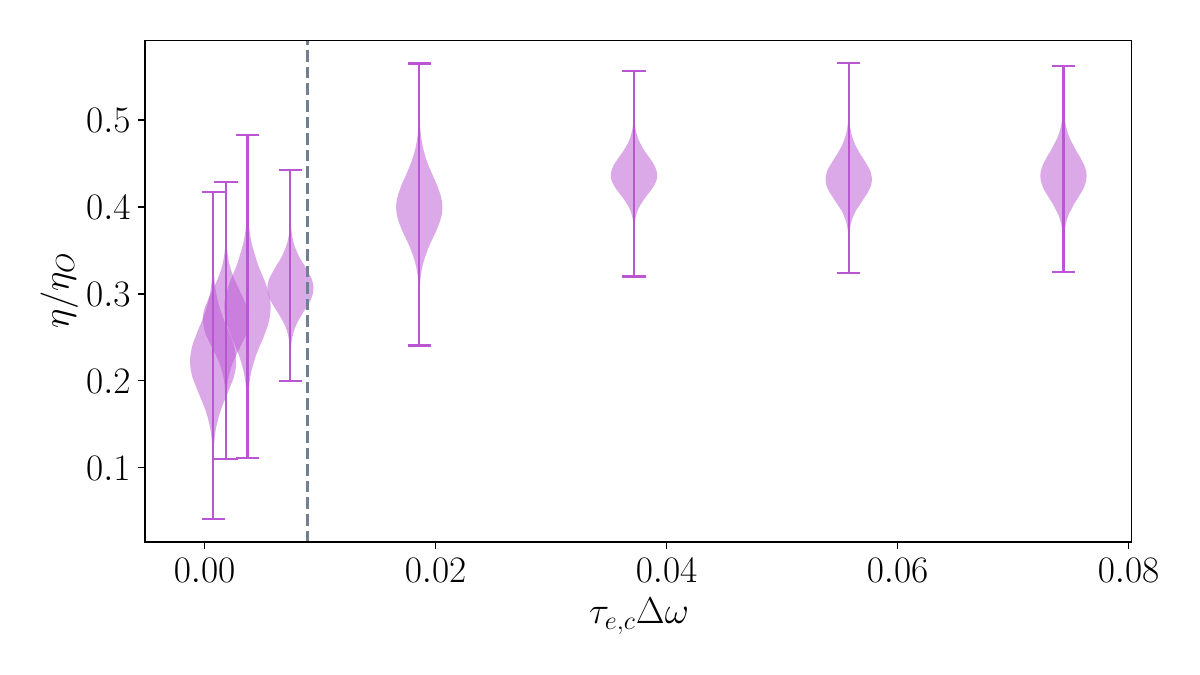}
    \caption{Efficiency of the cycles in units of the ideal Otto efficiency $\eta_O$, as a function of $\tau_{e,c}$, for the parameters listed in Sec.~\ref{sec:numerics}. The shaded areas indicate the PDFs of the efficiencies, and the error bars indicate the minimum and maximum efficiencies obtained. The vertical dashed line indicates the time $\omega_c^{-1}$. In a non-interacting condensate, the expansion and contraction times must be much larger than this value to achieve adiabaticity.}
    \label{fig:efficiency vs time}
\end{figure}

Let us first consider the impact of the speed of the adiabatic stroke. To this end,  for the set of parameters introduced in Sec.~\ref{sec:numerics}, we  performed several cycles with adiabatic strokes of different lengths $\tau_{e,c}$ (longer $\tau_{e,c}$ corresponds to slower expansions and contractions). 
In this case, we expect to attain the utmost efficiency for grater values of $\tau_{e,c}$, since the dynamics gets closer to the adiabatic limit. Figure \ref{fig:efficiency vs time} shows the efficiency distribution of the cycles for different values of $\tau_{e,c}$ (in units of $\Delta \omega^{-1}$). As expected, we find that the mean efficiency grows with this time and, for times $\tau_{e,c}$ much longer than the characteristic time associated to the adiabatic limit for non-interacting gases ($\sim\omega_h^{-1})$, the efficiency reaches a value that is independent of $\tau_{e,c}$. In this regime, the efficiency is roughly half that of the ideal Otto efficiency for a non-interacting gas.

\begin{figure}[tb]
    \centering
    \includegraphics[width=\columnwidth]{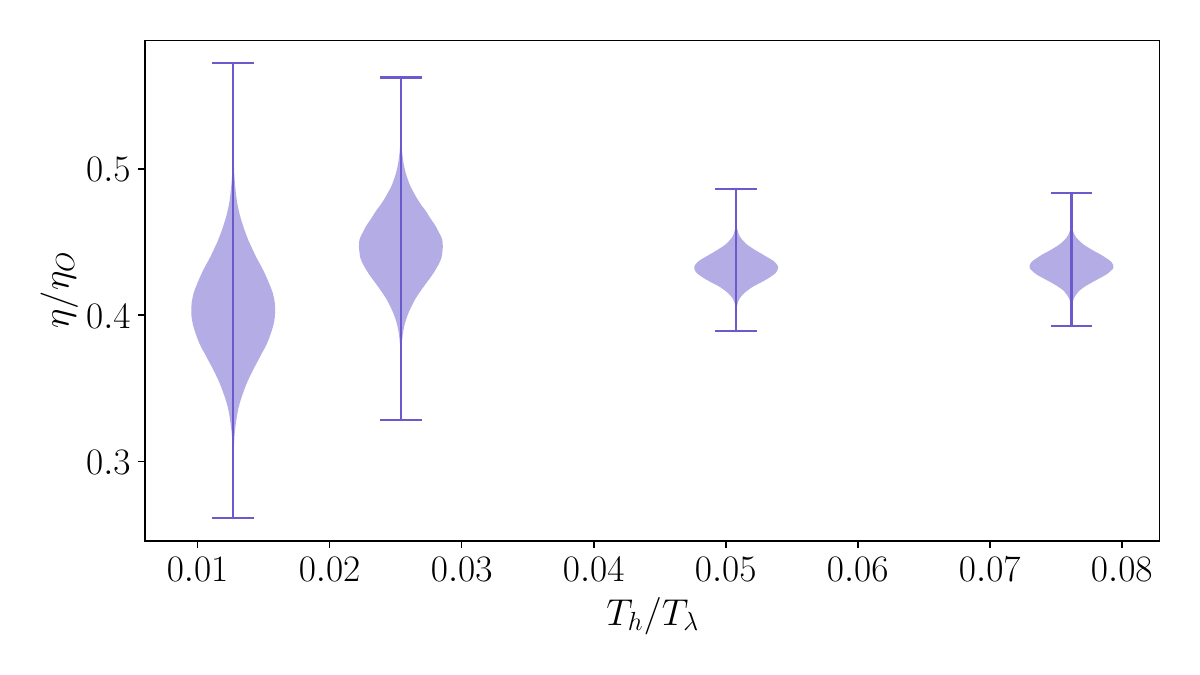}
    \caption{Efficiency (in units of the ideal Otto efficiency) as a function of the hot temperature $T_h$ using parameters the same parameters of Sec.~\ref{sec:numerics}. Labels for the markers are the same as in Fig.~\ref{fig:efficiency vs time}.}
    \label{fig:effs vs temp}
\end{figure}

 On the other hand, when only the temperature of the hot reservoir, $T_h$, is varied we find that the efficiency remains approximately constant (i.e., within error bars, see Fig.~\ref{fig:effs vs temp}). However, increasing $T_h$ (and therefore, the difference between $T_h$ and $T_c$) results in a reduction of fluctuations. Thus, larger temperature gradients leads to a better determination of the averaged efficiency. Note that this is also what happens with the ideal Otto cycle, where the efficiency is independent of the temperatures.

We will now analyse the efficiency in terms of the interaction strength. In this case, we expect that in the limit of a non-interacting gas the efficiency should approach $\eta_O$ as defined in Eq.~\eqref{eq:etaO}~\cite{myers2022boosting}. However, our method can only attain this limit asymptotically. This is due to the fact that when the interaction is removed, $g=0$, the thermalization time extends to infinity (as illustrated in Fig.~\ref{fig:cycle plot}). Therefore, we will evaluate the efficiency as the interaction strength is reduced. 
Reducing the interaction strength leads to a decrease of the speed of sound, and an increase in the healing length (i.e., a more dilute gas).
In the following, we express the results with respect to a coefficient $\alpha$ defined as
\begin{equation}
    g = \alpha g_0,
\end{equation}
where $0 < \alpha \le 1$, and $g_0$ corresponds to setting the speed of sound $c = (g_0 \rho_0/m)^{1/2} = 1U$ and $\xi = \hbar/(2m \rho_0 g_0)^{1/2} = 0.0707 L$ (i.e., the value used so far in this work).

\begin{figure}[tb]
    \centering
    \includegraphics[width=\columnwidth]{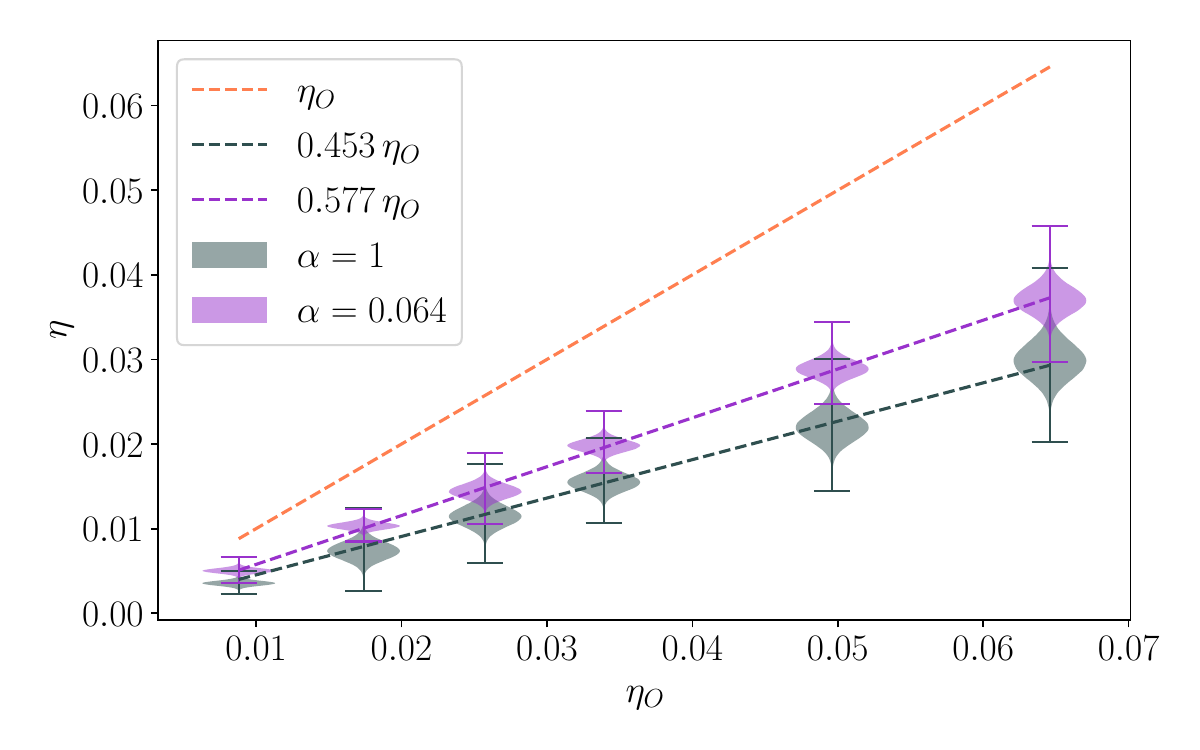}
    \caption{Efficiency of the cycle $\eta$ as a function of $\eta_O=1-\omega_c/\omega_h$, as we vary $\omega_c/\omega_h$ (symbols with the same color). For an ideal Otto cycle we expect $\eta = \eta_O$. Different colors of the symbols correspond to different interaction strengths: gray for $\alpha=1$ ($g=g_0$) and purple for $\alpha=0.064$ ($g=0.064 g_0$). Two slopes are indicated as references.}
    \label{fig:effs vs omega}
\end{figure}

First, we performed several cycles with different values of $\omega_c/\omega_h$, varying $\omega_c$ for two interaction strengths $\alpha=1$ and $\alpha = 0.064$
(the other parameters are the same as in Sec.~\ref{sec:numerics}).

Fig. \ref{fig:effs vs omega} shows the efficiency as a function of the ideal Otto efficiency $\eta_O = 1 - \omega_c/\omega_h$. We can see that it remains smaller than the Otto efficiency (which is indicated as a reference by a black dashed line). However, it still scales linearly with $1 - \omega_c/\omega_h$, and also gets closer to $\eta_O$ when $\alpha$ decreases. 
Interestingly, the behavior in Figs.~\ref{fig:effs vs temp}  and~\ref{fig:effs vs omega} indicate that the efficiency is independent of the temperature and the dependence with $\alpha$ can be factorized. This, at least in the regime of parameters that we are exploring, suggests that efficiency for the interacting gas is proportional to the ideal non-interacting Otto efficiency, with a proportionality factor that decreases with the interaction strength.

\begin{figure}[tb]
    \centering
    \includegraphics[width=\columnwidth]{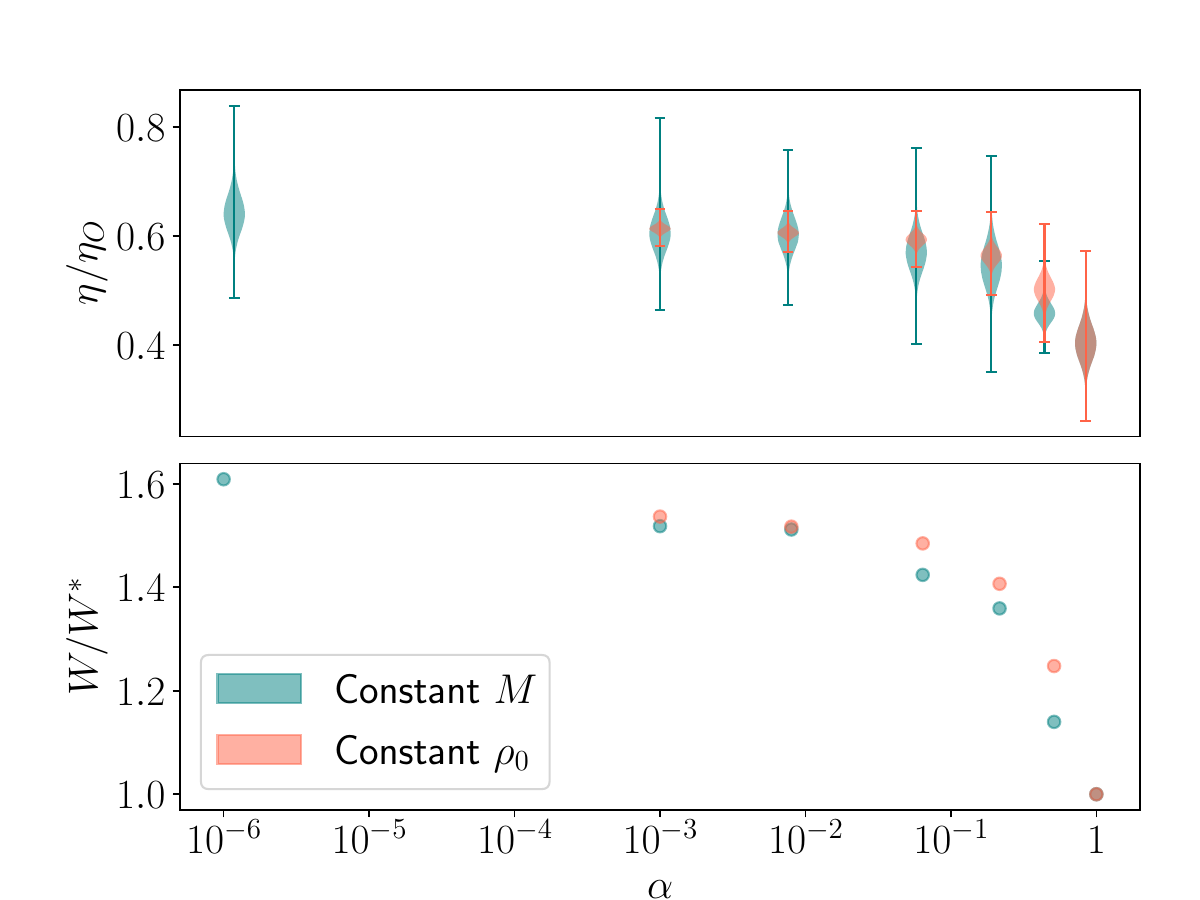}
\caption{Top: Efficiency in units of $\eta_O$ for different interaction strengths $\alpha = g/g_0$. We compare situations in which the  total mass in the condensate is constant (labeled as $M$), and in which the density in the center of the trap is constant (labeled as $\rho_0$). Labels for the markers are the same as in Fig.~\ref{fig:efficiency vs time}. Bottom: Mean work extracted by the engine as a function of $\alpha$, in units of the work $W^*$ for the fully interacting case with $\alpha=1$.}
    \label{fig:efficiency and power }
\end{figure}

Then, we fix the value of $\omega_c/\omega_h$ and vary the interaction strength.
Note that the volume of the condensate depends on $g$. In this case, as we consider repulsive interactions, the volume decreases with $g$  at a fixed potential. We considered two cases different situations: one in which the total mass of the condensate is kept constant as the interaction strength is changed, and the other in which the density in the center of the trap is kept constant.  We can appreciate from the top panel of Fig.~\ref{fig:efficiency and power } that both cases display similar efficiencies. However, when the mass is constant the fluctuations are larger than when the density is kept constant. This stems from the fact that as the interaction strength is reduced, the concentration of particles at the center of the trap decrease substantially, thus increasing the amount of  fluctuations. In general, we observe that the efficiency grows slowly for small $\alpha$. In both cases, as $\alpha$ decreases the efficiency increases attaining a mean value $\approx 65 \%$ of $\eta_O$ for the minimum $\alpha$ considered in the simulations.
As previously stated, we are unable to reduce the value of $\alpha$ any further as calculating isochoric strokes becomes excessively expensive with a progressively increasing thermalisation time.

Finally, we analyze the power of this engine. 
Let us first look at the bottom panel of Fig.~\ref{fig:efficiency and power } where the mean extracted work as a function of the interaction strength $\alpha$ is shown. In this case, we compare the extracted amount of work for a given value of $\alpha$ with the work $W^*$ extracted in the fully interacting case with $\alpha = 1$. Note that $W/W^*$ increases by $\approx 50 \%$ for decreasing $\alpha$, and becomes approximately constant for $\alpha < 10^{-1}$. The actual power of the cycle is determined by the ratio of the work to the time required to complete the cycle. As it occurs in both  numerical simulations and in a real gases, we consider that the   thermalization times in the isochoric strokes are longer than the times required for the expansion and compression. Thus, we can approximate the length of the cycle as twice the length of the thermalization process.  
Note that, in our simulations we use the stochastic Ginzburg-Landau equation as a multivariate Fokker-Plank equation to obtain the new equilibria (at a given temperature) of the Grand canonical ensemble, therefore the time in the simulation should not be directly associated to an actual thermalization time. However, in the non-interacting limit the thermalization time effectively goes to infinity, as the time between collisions diverges. 

We can still estimate the order of the thermalization  time for the interacting case from kinetic theory. Note that $g \sim a $, i.e., it is linearly proportional to the scattering length, and thus $g \sim a \sim \sqrt{\sigma}$ where $\sigma$ is the collision cross section. As $\sigma \sim 1/\tau$ where $\tau$ is the time between collisions, for a fixed number of particles the time it takes for the system to thermalize with $\alpha=1$ compared with the time when $\alpha < 1$ is proportional to the ratio of the times between collisions,
\begin{equation}
    \frac{\tau_0}{\tau} \sim \left( \frac{g}{g_0} \right)^2 = \alpha^2,
\end{equation}
where $\tau_0$ is the value of $\tau$ when $\alpha=1$. 
This indicates (in qualitative agreement with the results from the numerical simulations) that interactions allow for much faster cycles, and extraction of significantly more power (e.g., from the cycles in the plateau of $W$ for $\alpha < 10^{-1}$ in Fig.~\ref{fig:efficiency and power }, even with the reduction of the extracted work of $\approx 50 \%$ with respect to $\alpha = 1$). In other words, interacting gases allow to get higher power in a finite time cycle.
Moreover, in principle by adjusting the interaction strength of the condensate, a power enhancement at nearly constant efficiency can be achieved (see Figs.~\ref{fig:effs vs omega} and~\ref{fig:efficiency and power }).

\section{Conclusions}

In this work, we performed numerical simulations of quantum Otto engines that have an interacting BEC as its working medium. We were able to recover not only the thermodynamics of the system, but also its complete dynamics, which enable in turn to perform a  detailed analysis of the engine. Analyzing, for instance, the different contributions to the energy along the adiabatic strokes. 

We characterized the efficiency of the engine by performing several simulations in which we independently changed the temperatures, the trap frequencies, and the interaction strength of the gas. We found that the efficiency is independent of the temperature. However, fluctuations in the efficiency and in other observables are reduced as the difference between the temperatures of the reservoirs increases. Also, their dependence on the trap frequencies turns out to be similar to the non-interacting case, but with a proportionality factor that depends on the interaction strength. 

We also show that the efficiency and work output of the engine decrease as the interaction strength of the BEC becomes larger. However, the timescale it takes to the system to thermalize is inversely proportional to the square of the interaction strength. Thus, for small interactions, we find a regime in which increasing the interaction of the BEC allows for a considerable increase in power, while the efficiency is only slightly reduced. Since the interaction strength of the BEC can be experimentally tuned, our results provide a possible way to improve the power of a quantum engine at a small cost in efficiency.

\begin{acknowledgments}
J.A.E.~and F.M.~thank Muriel Bonetto and Facundo Pugliese for useful discussions and suggestions. J.A.E.~and P.D.M. acknowledge financial support from UBACyT Grant No.~20020170100508BA and PICT Grant No.~2018-4298. F.M.~and~A.J.R acknowledge financial support from  PICT Grant No.~2019-4349 and No.~2021-01288.
\end{acknowledgments}

\bibliography{ms}

\begin{thebibliography}{41}%
\makeatletter
\providecommand \@ifxundefined [1]{%
 \@ifx{#1\undefined}
}%
\providecommand \@ifnum [1]{%
 \ifnum #1\expandafter \@firstoftwo
 \else \expandafter \@secondoftwo
 \fi
}%
\providecommand \@ifx [1]{%
 \ifx #1\expandafter \@firstoftwo
 \else \expandafter \@secondoftwo
 \fi
}%
\providecommand \natexlab [1]{#1}%
\providecommand \enquote  [1]{``#1''}%
\providecommand \bibnamefont  [1]{#1}%
\providecommand \bibfnamefont [1]{#1}%
\providecommand \citenamefont [1]{#1}%
\providecommand \href@noop [0]{\@secondoftwo}%
\providecommand \href [0]{\begingroup \@sanitize@url \@href}%
\providecommand \@href[1]{\@@startlink{#1}\@@href}%
\providecommand \@@href[1]{\endgroup#1\@@endlink}%
\providecommand \@sanitize@url [0]{\catcode `\\12\catcode `\$12\catcode
  `\&12\catcode `\#12\catcode `\^12\catcode `\_12\catcode `\%12\relax}%
\providecommand \@@startlink[1]{}%
\providecommand \@@endlink[0]{}%
\providecommand \url  [0]{\begingroup\@sanitize@url \@url }%
\providecommand \@url [1]{\endgroup\@href {#1}{\urlprefix }}%
\providecommand \urlprefix  [0]{URL }%
\providecommand \Eprint [0]{\href }%
\providecommand \doibase [0]{http://dx.doi.org/}%
\providecommand \selectlanguage [0]{\@gobble}%
\providecommand \bibinfo  [0]{\@secondoftwo}%
\providecommand \bibfield  [0]{\@secondoftwo}%
\providecommand \translation [1]{[#1]}%
\providecommand \BibitemOpen [0]{}%
\providecommand \bibitemStop [0]{}%
\providecommand \bibitemNoStop [0]{.\EOS\space}%
\providecommand \EOS [0]{\spacefactor3000\relax}%
\providecommand \BibitemShut  [1]{\csname bibitem#1\endcsname}%
\let\auto@bib@innerbib\@empty
\bibitem [{\citenamefont {Vinjanampathy}\ and\ \citenamefont
  {Anders}(2016)}]{vinjanampathy2016quantum}%
  \BibitemOpen
  \bibfield  {author} {\bibinfo {author} {\bibfnamefont {S.}~\bibnamefont
  {Vinjanampathy}}\ and\ \bibinfo {author} {\bibfnamefont {J.}~\bibnamefont
  {Anders}},\ }\href@noop {} {\bibfield  {journal} {\bibinfo  {journal}
  {Contemporary Physics}\ }\textbf {\bibinfo {volume} {57}},\ \bibinfo {pages}
  {545} (\bibinfo {year} {2016})}\BibitemShut {NoStop}%
\bibitem [{\citenamefont {Binder}\ \emph {et~al.}(2018)\citenamefont {Binder},
  \citenamefont {Correa}, \citenamefont {Gogolin}, \citenamefont {Anders},\
  and\ \citenamefont {Adesso}}]{binder2018thermodynamics}%
  \BibitemOpen
  \bibfield  {author} {\bibinfo {author} {\bibfnamefont {F.}~\bibnamefont
  {Binder}}, \bibinfo {author} {\bibfnamefont {L.~A.}\ \bibnamefont {Correa}},
  \bibinfo {author} {\bibfnamefont {C.}~\bibnamefont {Gogolin}}, \bibinfo
  {author} {\bibfnamefont {J.}~\bibnamefont {Anders}}, \ and\ \bibinfo {author}
  {\bibfnamefont {G.}~\bibnamefont {Adesso}},\ }\href@noop {} {\bibfield
  {journal} {\bibinfo  {journal} {Fundamental Theories of Physics}\ }\textbf
  {\bibinfo {volume} {195}},\ \bibinfo {pages} {1} (\bibinfo {year}
  {2018})}\BibitemShut {NoStop}%
\bibitem [{\citenamefont {Myers}\ \emph
  {et~al.}(2022{\natexlab{a}})\citenamefont {Myers}, \citenamefont {Abah},\
  and\ \citenamefont {Deffner}}]{myers2022quantum}%
  \BibitemOpen
  \bibfield  {author} {\bibinfo {author} {\bibfnamefont {N.~M.}\ \bibnamefont
  {Myers}}, \bibinfo {author} {\bibfnamefont {O.}~\bibnamefont {Abah}}, \ and\
  \bibinfo {author} {\bibfnamefont {S.}~\bibnamefont {Deffner}},\ }\href@noop
  {} {\bibfield  {journal} {\bibinfo  {journal} {AVS Quantum Science}\ }\textbf
  {\bibinfo {volume} {4}} (\bibinfo {year} {2022}{\natexlab{a}})}\BibitemShut
  {NoStop}%
\bibitem [{\citenamefont {Mitchison}(2019)}]{mitchison2019quantum}%
  \BibitemOpen
  \bibfield  {author} {\bibinfo {author} {\bibfnamefont {M.~T.}\ \bibnamefont
  {Mitchison}},\ }\href@noop {} {\bibfield  {journal} {\bibinfo  {journal}
  {Contemporary Physics}\ }\textbf {\bibinfo {volume} {60}},\ \bibinfo {pages}
  {164} (\bibinfo {year} {2019})}\BibitemShut {NoStop}%
\bibitem [{\citenamefont {Bhattacharjee}\ and\ \citenamefont
  {Dutta}(2021)}]{bhattacharjee2021quantum}%
  \BibitemOpen
  \bibfield  {author} {\bibinfo {author} {\bibfnamefont {S.}~\bibnamefont
  {Bhattacharjee}}\ and\ \bibinfo {author} {\bibfnamefont {A.}~\bibnamefont
  {Dutta}},\ }\href@noop {} {\bibfield  {journal} {\bibinfo  {journal} {The
  European Physical Journal B}\ }\textbf {\bibinfo {volume} {94}},\ \bibinfo
  {pages} {1} (\bibinfo {year} {2021})}\BibitemShut {NoStop}%
\bibitem [{\citenamefont {Fialko}\ and\ \citenamefont
  {Hallwood}(2012)}]{fialko2012isolated}%
  \BibitemOpen
  \bibfield  {author} {\bibinfo {author} {\bibfnamefont {O.}~\bibnamefont
  {Fialko}}\ and\ \bibinfo {author} {\bibfnamefont {D.}~\bibnamefont
  {Hallwood}},\ }\href@noop {} {\bibfield  {journal} {\bibinfo  {journal}
  {Physical review letters}\ }\textbf {\bibinfo {volume} {108}},\ \bibinfo
  {pages} {085303} (\bibinfo {year} {2012})}\BibitemShut {NoStop}%
\bibitem [{\citenamefont {Camati}\ \emph {et~al.}(2019)\citenamefont {Camati},
  \citenamefont {Santos},\ and\ \citenamefont {Serra}}]{camati2019coherence}%
  \BibitemOpen
  \bibfield  {author} {\bibinfo {author} {\bibfnamefont {P.~A.}\ \bibnamefont
  {Camati}}, \bibinfo {author} {\bibfnamefont {J.~F.}\ \bibnamefont {Santos}},
  \ and\ \bibinfo {author} {\bibfnamefont {R.~M.}\ \bibnamefont {Serra}},\
  }\href@noop {} {\bibfield  {journal} {\bibinfo  {journal} {Physical Review
  A}\ }\textbf {\bibinfo {volume} {99}},\ \bibinfo {pages} {062103} (\bibinfo
  {year} {2019})}\BibitemShut {NoStop}%
\bibitem [{\citenamefont {Dann}\ and\ \citenamefont
  {Kosloff}(2020)}]{dann2020quantum}%
  \BibitemOpen
  \bibfield  {author} {\bibinfo {author} {\bibfnamefont {R.}~\bibnamefont
  {Dann}}\ and\ \bibinfo {author} {\bibfnamefont {R.}~\bibnamefont {Kosloff}},\
  }\href@noop {} {\bibfield  {journal} {\bibinfo  {journal} {New Journal of
  Physics}\ }\textbf {\bibinfo {volume} {22}},\ \bibinfo {pages} {013055}
  (\bibinfo {year} {2020})}\BibitemShut {NoStop}%
\bibitem [{\citenamefont {Hewgill}\ \emph {et~al.}(2018)\citenamefont
  {Hewgill}, \citenamefont {Ferraro},\ and\ \citenamefont
  {De~Chiara}}]{hewgill2018quantum}%
  \BibitemOpen
  \bibfield  {author} {\bibinfo {author} {\bibfnamefont {A.}~\bibnamefont
  {Hewgill}}, \bibinfo {author} {\bibfnamefont {A.}~\bibnamefont {Ferraro}}, \
  and\ \bibinfo {author} {\bibfnamefont {G.}~\bibnamefont {De~Chiara}},\
  }\href@noop {} {\bibfield  {journal} {\bibinfo  {journal} {Physical Review
  A}\ }\textbf {\bibinfo {volume} {98}},\ \bibinfo {pages} {042102} (\bibinfo
  {year} {2018})}\BibitemShut {NoStop}%
\bibitem [{\citenamefont {Elouard}\ \emph {et~al.}(2017)\citenamefont
  {Elouard}, \citenamefont {Herrera-Mart{\'\i}}, \citenamefont {Huard},\ and\
  \citenamefont {Auffeves}}]{elouard2017extracting}%
  \BibitemOpen
  \bibfield  {author} {\bibinfo {author} {\bibfnamefont {C.}~\bibnamefont
  {Elouard}}, \bibinfo {author} {\bibfnamefont {D.}~\bibnamefont
  {Herrera-Mart{\'\i}}}, \bibinfo {author} {\bibfnamefont {B.}~\bibnamefont
  {Huard}}, \ and\ \bibinfo {author} {\bibfnamefont {A.}~\bibnamefont
  {Auffeves}},\ }\href@noop {} {\bibfield  {journal} {\bibinfo  {journal}
  {Physical Review Letters}\ }\textbf {\bibinfo {volume} {118}},\ \bibinfo
  {pages} {260603} (\bibinfo {year} {2017})}\BibitemShut {NoStop}%
\bibitem [{\citenamefont {Elouard}\ and\ \citenamefont
  {Jordan}(2018)}]{elouard2018efficient}%
  \BibitemOpen
  \bibfield  {author} {\bibinfo {author} {\bibfnamefont {C.}~\bibnamefont
  {Elouard}}\ and\ \bibinfo {author} {\bibfnamefont {A.~N.}\ \bibnamefont
  {Jordan}},\ }\href@noop {} {\bibfield  {journal} {\bibinfo  {journal}
  {Physical Review Letters}\ }\textbf {\bibinfo {volume} {120}},\ \bibinfo
  {pages} {260601} (\bibinfo {year} {2018})}\BibitemShut {NoStop}%
\bibitem [{\citenamefont {Jordan}\ \emph {et~al.}(2020)\citenamefont {Jordan},
  \citenamefont {Elouard},\ and\ \citenamefont
  {Auff{\`e}ves}}]{jordan2020quantum}%
  \BibitemOpen
  \bibfield  {author} {\bibinfo {author} {\bibfnamefont {A.~N.}\ \bibnamefont
  {Jordan}}, \bibinfo {author} {\bibfnamefont {C.}~\bibnamefont {Elouard}}, \
  and\ \bibinfo {author} {\bibfnamefont {A.}~\bibnamefont {Auff{\`e}ves}},\
  }\href@noop {} {\bibfield  {journal} {\bibinfo  {journal} {Quantum Studies:
  Mathematics and Foundations}\ }\textbf {\bibinfo {volume} {7}},\ \bibinfo
  {pages} {203} (\bibinfo {year} {2020})}\BibitemShut {NoStop}%
\bibitem [{\citenamefont {Ro{\ss}nagel}\ \emph {et~al.}(2016)\citenamefont
  {Ro{\ss}nagel}, \citenamefont {Dawkins}, \citenamefont {Tolazzi},
  \citenamefont {Abah}, \citenamefont {Lutz}, \citenamefont {Schmidt-Kaler},\
  and\ \citenamefont {Singer}}]{rossnagel2016single}%
  \BibitemOpen
  \bibfield  {author} {\bibinfo {author} {\bibfnamefont {J.}~\bibnamefont
  {Ro{\ss}nagel}}, \bibinfo {author} {\bibfnamefont {S.~T.}\ \bibnamefont
  {Dawkins}}, \bibinfo {author} {\bibfnamefont {K.~N.}\ \bibnamefont
  {Tolazzi}}, \bibinfo {author} {\bibfnamefont {O.}~\bibnamefont {Abah}},
  \bibinfo {author} {\bibfnamefont {E.}~\bibnamefont {Lutz}}, \bibinfo {author}
  {\bibfnamefont {F.}~\bibnamefont {Schmidt-Kaler}}, \ and\ \bibinfo {author}
  {\bibfnamefont {K.}~\bibnamefont {Singer}},\ }\href@noop {} {\bibfield
  {journal} {\bibinfo  {journal} {Science}\ }\textbf {\bibinfo {volume}
  {352}},\ \bibinfo {pages} {325} (\bibinfo {year} {2016})}\BibitemShut
  {NoStop}%
\bibitem [{\citenamefont {Abah}\ \emph {et~al.}(2012)\citenamefont {Abah},
  \citenamefont {Rossnagel}, \citenamefont {Jacob}, \citenamefont {Deffner},
  \citenamefont {Schmidt-Kaler}, \citenamefont {Singer},\ and\ \citenamefont
  {Lutz}}]{abah2012single}%
  \BibitemOpen
  \bibfield  {author} {\bibinfo {author} {\bibfnamefont {O.}~\bibnamefont
  {Abah}}, \bibinfo {author} {\bibfnamefont {J.}~\bibnamefont {Rossnagel}},
  \bibinfo {author} {\bibfnamefont {G.}~\bibnamefont {Jacob}}, \bibinfo
  {author} {\bibfnamefont {S.}~\bibnamefont {Deffner}}, \bibinfo {author}
  {\bibfnamefont {F.}~\bibnamefont {Schmidt-Kaler}}, \bibinfo {author}
  {\bibfnamefont {K.}~\bibnamefont {Singer}}, \ and\ \bibinfo {author}
  {\bibfnamefont {E.}~\bibnamefont {Lutz}},\ }\href@noop {} {\bibfield
  {journal} {\bibinfo  {journal} {Physical Review Letters}\ }\textbf {\bibinfo
  {volume} {109}},\ \bibinfo {pages} {203006} (\bibinfo {year}
  {2012})}\BibitemShut {NoStop}%
\bibitem [{\citenamefont {Von~Lindenfels}\ \emph {et~al.}(2019)\citenamefont
  {Von~Lindenfels}, \citenamefont {Gr{\"a}b}, \citenamefont {Schmiegelow},
  \citenamefont {Kaushal}, \citenamefont {Schulz}, \citenamefont {Mitchison},
  \citenamefont {Goold}, \citenamefont {Schmidt-Kaler},\ and\ \citenamefont
  {Poschinger}}]{von2019spin}%
  \BibitemOpen
  \bibfield  {author} {\bibinfo {author} {\bibfnamefont {D.}~\bibnamefont
  {Von~Lindenfels}}, \bibinfo {author} {\bibfnamefont {O.}~\bibnamefont
  {Gr{\"a}b}}, \bibinfo {author} {\bibfnamefont {C.~T.}\ \bibnamefont
  {Schmiegelow}}, \bibinfo {author} {\bibfnamefont {V.}~\bibnamefont
  {Kaushal}}, \bibinfo {author} {\bibfnamefont {J.}~\bibnamefont {Schulz}},
  \bibinfo {author} {\bibfnamefont {M.~T.}\ \bibnamefont {Mitchison}}, \bibinfo
  {author} {\bibfnamefont {J.}~\bibnamefont {Goold}}, \bibinfo {author}
  {\bibfnamefont {F.}~\bibnamefont {Schmidt-Kaler}}, \ and\ \bibinfo {author}
  {\bibfnamefont {U.~G.}\ \bibnamefont {Poschinger}},\ }\href@noop {}
  {\bibfield  {journal} {\bibinfo  {journal} {Physical Review Letters}\
  }\textbf {\bibinfo {volume} {123}},\ \bibinfo {pages} {080602} (\bibinfo
  {year} {2019})}\BibitemShut {NoStop}%
\bibitem [{\citenamefont {Myers}\ \emph
  {et~al.}(2022{\natexlab{b}})\citenamefont {Myers}, \citenamefont {Pe{\~n}a},
  \citenamefont {Negrete}, \citenamefont {Vargas}, \citenamefont {De~Chiara},\
  and\ \citenamefont {Deffner}}]{myers2022boosting}%
  \BibitemOpen
  \bibfield  {author} {\bibinfo {author} {\bibfnamefont {N.~M.}\ \bibnamefont
  {Myers}}, \bibinfo {author} {\bibfnamefont {F.~J.}\ \bibnamefont {Pe{\~n}a}},
  \bibinfo {author} {\bibfnamefont {O.}~\bibnamefont {Negrete}}, \bibinfo
  {author} {\bibfnamefont {P.}~\bibnamefont {Vargas}}, \bibinfo {author}
  {\bibfnamefont {G.}~\bibnamefont {De~Chiara}}, \ and\ \bibinfo {author}
  {\bibfnamefont {S.}~\bibnamefont {Deffner}},\ }\href@noop {} {\bibfield
  {journal} {\bibinfo  {journal} {New Journal of Physics}\ }\textbf {\bibinfo
  {volume} {24}},\ \bibinfo {pages} {025001} (\bibinfo {year}
  {2022}{\natexlab{b}})}\BibitemShut {NoStop}%
\bibitem [{\citenamefont {Reyes-Ayala}\ \emph {et~al.}(2023)\citenamefont
  {Reyes-Ayala}, \citenamefont {Miotti}, \citenamefont {Hemmerling},
  \citenamefont {Dubessy}, \citenamefont {Perrin}, \citenamefont
  {Romero-Rochin},\ and\ \citenamefont {Bagnato}}]{reyes2023carnot}%
  \BibitemOpen
  \bibfield  {author} {\bibinfo {author} {\bibfnamefont {I.}~\bibnamefont
  {Reyes-Ayala}}, \bibinfo {author} {\bibfnamefont {M.}~\bibnamefont {Miotti}},
  \bibinfo {author} {\bibfnamefont {M.}~\bibnamefont {Hemmerling}}, \bibinfo
  {author} {\bibfnamefont {R.}~\bibnamefont {Dubessy}}, \bibinfo {author}
  {\bibfnamefont {H.}~\bibnamefont {Perrin}}, \bibinfo {author} {\bibfnamefont
  {V.}~\bibnamefont {Romero-Rochin}}, \ and\ \bibinfo {author} {\bibfnamefont
  {V.~S.}\ \bibnamefont {Bagnato}},\ }\href@noop {} {\bibfield  {journal}
  {\bibinfo  {journal} {Entropy}\ }\textbf {\bibinfo {volume} {25}},\ \bibinfo
  {pages} {311} (\bibinfo {year} {2023})}\BibitemShut {NoStop}%
\bibitem [{\citenamefont {Keller}\ \emph {et~al.}(2020)\citenamefont {Keller},
  \citenamefont {Fogarty}, \citenamefont {Li},\ and\ \citenamefont
  {Busch}}]{keller2020feshbach}%
  \BibitemOpen
  \bibfield  {author} {\bibinfo {author} {\bibfnamefont {T.}~\bibnamefont
  {Keller}}, \bibinfo {author} {\bibfnamefont {T.}~\bibnamefont {Fogarty}},
  \bibinfo {author} {\bibfnamefont {J.}~\bibnamefont {Li}}, \ and\ \bibinfo
  {author} {\bibfnamefont {T.}~\bibnamefont {Busch}},\ }\href@noop {}
  {\bibfield  {journal} {\bibinfo  {journal} {Physical Review Research}\
  }\textbf {\bibinfo {volume} {2}},\ \bibinfo {pages} {033335} (\bibinfo {year}
  {2020})}\BibitemShut {NoStop}%
\bibitem [{\citenamefont {Li}\ \emph {et~al.}(2018)\citenamefont {Li},
  \citenamefont {Fogarty}, \citenamefont {Campbell}, \citenamefont {Chen},\
  and\ \citenamefont {Busch}}]{li2018efficient}%
  \BibitemOpen
  \bibfield  {author} {\bibinfo {author} {\bibfnamefont {J.}~\bibnamefont
  {Li}}, \bibinfo {author} {\bibfnamefont {T.}~\bibnamefont {Fogarty}},
  \bibinfo {author} {\bibfnamefont {S.}~\bibnamefont {Campbell}}, \bibinfo
  {author} {\bibfnamefont {X.}~\bibnamefont {Chen}}, \ and\ \bibinfo {author}
  {\bibfnamefont {T.}~\bibnamefont {Busch}},\ }\href@noop {} {\bibfield
  {journal} {\bibinfo  {journal} {New Journal of Physics}\ }\textbf {\bibinfo
  {volume} {20}},\ \bibinfo {pages} {015005} (\bibinfo {year}
  {2018})}\BibitemShut {NoStop}%
\bibitem [{\citenamefont {Niedenzu}\ \emph {et~al.}(2019)\citenamefont
  {Niedenzu}, \citenamefont {Mazets}, \citenamefont {Kurizki},\ and\
  \citenamefont {Jendrzejewski}}]{niedenzu2019quantized}%
  \BibitemOpen
  \bibfield  {author} {\bibinfo {author} {\bibfnamefont {W.}~\bibnamefont
  {Niedenzu}}, \bibinfo {author} {\bibfnamefont {I.}~\bibnamefont {Mazets}},
  \bibinfo {author} {\bibfnamefont {G.}~\bibnamefont {Kurizki}}, \ and\
  \bibinfo {author} {\bibfnamefont {F.}~\bibnamefont {Jendrzejewski}},\
  }\href@noop {} {\bibfield  {journal} {\bibinfo  {journal} {Quantum}\ }\textbf
  {\bibinfo {volume} {3}},\ \bibinfo {pages} {155} (\bibinfo {year}
  {2019})}\BibitemShut {NoStop}%
\bibitem [{\citenamefont {Gluza}\ \emph {et~al.}(2021)\citenamefont {Gluza},
  \citenamefont {Sabino}, \citenamefont {Ng}, \citenamefont {Vitagliano},
  \citenamefont {Pezzutto}, \citenamefont {Omar}, \citenamefont {Mazets},
  \citenamefont {Huber}, \citenamefont {Schmiedmayer},\ and\ \citenamefont
  {Eisert}}]{gluza2021quantum}%
  \BibitemOpen
  \bibfield  {author} {\bibinfo {author} {\bibfnamefont {M.}~\bibnamefont
  {Gluza}}, \bibinfo {author} {\bibfnamefont {J.}~\bibnamefont {Sabino}},
  \bibinfo {author} {\bibfnamefont {N.~H.}\ \bibnamefont {Ng}}, \bibinfo
  {author} {\bibfnamefont {G.}~\bibnamefont {Vitagliano}}, \bibinfo {author}
  {\bibfnamefont {M.}~\bibnamefont {Pezzutto}}, \bibinfo {author}
  {\bibfnamefont {Y.}~\bibnamefont {Omar}}, \bibinfo {author} {\bibfnamefont
  {I.}~\bibnamefont {Mazets}}, \bibinfo {author} {\bibfnamefont
  {M.}~\bibnamefont {Huber}}, \bibinfo {author} {\bibfnamefont
  {J.}~\bibnamefont {Schmiedmayer}}, \ and\ \bibinfo {author} {\bibfnamefont
  {J.}~\bibnamefont {Eisert}},\ }\href@noop {} {\bibfield  {journal} {\bibinfo
  {journal} {PRX Quantum}\ }\textbf {\bibinfo {volume} {2}},\ \bibinfo {pages}
  {030310} (\bibinfo {year} {2021})}\BibitemShut {NoStop}%
\bibitem [{\citenamefont {Krstulovic}\ and\ \citenamefont
  {Brachet}(2011)}]{Krstulovic2011}%
  \BibitemOpen
  \bibfield  {author} {\bibinfo {author} {\bibfnamefont {G.}~\bibnamefont
  {Krstulovic}}\ and\ \bibinfo {author} {\bibfnamefont {M.}~\bibnamefont
  {Brachet}},\ }\href {\doibase 10.1103/physreve.83.066311} {\bibfield
  {journal} {\bibinfo  {journal} {Physical Review E}\ }\textbf {\bibinfo
  {volume} {83}} (\bibinfo {year} {2011}),\
  10.1103/physreve.83.066311}\BibitemShut {NoStop}%
\bibitem [{\citenamefont {Estrada}\ \emph
  {et~al.}(2022{\natexlab{a}})\citenamefont {Estrada}, \citenamefont
  {Brachet},\ and\ \citenamefont {Mininni}}]{AmetteEstrada2022b}%
  \BibitemOpen
  \bibfield  {author} {\bibinfo {author} {\bibfnamefont {J.~A.}\ \bibnamefont
  {Estrada}}, \bibinfo {author} {\bibfnamefont {M.~E.}\ \bibnamefont
  {Brachet}}, \ and\ \bibinfo {author} {\bibfnamefont {P.~D.}\ \bibnamefont
  {Mininni}},\ }\href {\doibase 10.1116/5.0123277} {\bibfield  {journal}
  {\bibinfo  {journal} {{AVS} Quantum Science}\ }\textbf {\bibinfo {volume}
  {4}},\ \bibinfo {pages} {046201} (\bibinfo {year}
  {2022}{\natexlab{a}})}\BibitemShut {NoStop}%
\bibitem [{\citenamefont {Nore}\ \emph {et~al.}(1997)\citenamefont {Nore},
  \citenamefont {Abid},\ and\ \citenamefont {Brachet}}]{Nore1997}%
  \BibitemOpen
  \bibfield  {author} {\bibinfo {author} {\bibfnamefont {C.}~\bibnamefont
  {Nore}}, \bibinfo {author} {\bibfnamefont {M.}~\bibnamefont {Abid}}, \ and\
  \bibinfo {author} {\bibfnamefont {M.~E.}\ \bibnamefont {Brachet}},\ }\href
  {\doibase 10.1063/1.869473} {\bibfield  {journal} {\bibinfo  {journal}
  {Physics of Fluids}\ }\textbf {\bibinfo {volume} {9}},\ \bibinfo {pages}
  {2644} (\bibinfo {year} {1997})}\BibitemShut {NoStop}%
\bibitem [{\citenamefont {Severino}\ \emph {et~al.}(2022)\citenamefont
  {Severino}, \citenamefont {Mininni}, \citenamefont {Fradkin}, \citenamefont
  {Bekeris}, \citenamefont {Pasquini}, ,\ and\ \citenamefont
  {Lozano}}]{Severino_2022}%
  \BibitemOpen
  \bibfield  {author} {\bibinfo {author} {\bibfnamefont {R.~S.}\ \bibnamefont
  {Severino}}, \bibinfo {author} {\bibfnamefont {P.~D.}\ \bibnamefont
  {Mininni}}, \bibinfo {author} {\bibfnamefont {E.}~\bibnamefont {Fradkin}},
  \bibinfo {author} {\bibfnamefont {V.}~\bibnamefont {Bekeris}}, \bibinfo
  {author} {\bibfnamefont {G.}~\bibnamefont {Pasquini}}, , \ and\ \bibinfo
  {author} {\bibfnamefont {G.~S.}\ \bibnamefont {Lozano}},\ }\href@noop {}
  {\bibfield  {journal} {\bibinfo  {journal} {Physical Review B}\ }\textbf
  {\bibinfo {volume} {106}},\ \bibinfo {pages} {094512} (\bibinfo {year}
  {2022})}\BibitemShut {NoStop}%
\bibitem [{\citenamefont {{Van Kampen}}(2007)}]{VanKampen}%
  \BibitemOpen
  \bibfield  {author} {\bibinfo {author} {\bibfnamefont {N.}~\bibnamefont {{Van
  Kampen}}},\ }\href@noop {} {\emph {\bibinfo {title} {Stochastic Processes in
  Physics and Chemistry (Third Edition)}}}\ (\bibinfo  {publisher} {Elsevier},\
  \bibinfo {address} {Amsterdam},\ \bibinfo {year} {2007})\BibitemShut
  {NoStop}%
\bibitem [{\citenamefont {Gardiner}\ \emph {et~al.}(2002)\citenamefont
  {Gardiner}, \citenamefont {Anglin},\ and\ \citenamefont
  {Fudge}}]{Gardiner_2002}%
  \BibitemOpen
  \bibfield  {author} {\bibinfo {author} {\bibfnamefont {C.}~\bibnamefont
  {Gardiner}}, \bibinfo {author} {\bibfnamefont {J.}~\bibnamefont {Anglin}}, \
  and\ \bibinfo {author} {\bibfnamefont {T.}~\bibnamefont {Fudge}},\
  }\href@noop {} {\bibfield  {journal} {\bibinfo  {journal} {Journal of Physics
  B: Atomic, Molecular and Optical Physics}\ }\textbf {\bibinfo {volume}
  {35}},\ \bibinfo {pages} {1555} (\bibinfo {year} {2002})}\BibitemShut
  {NoStop}%
\bibitem [{\citenamefont {Calzetta}\ \emph {et~al.}(2007)\citenamefont
  {Calzetta}, \citenamefont {Hu},\ and\ \citenamefont
  {Verdaguer}}]{Calzetta_2007}%
  \BibitemOpen
  \bibfield  {author} {\bibinfo {author} {\bibfnamefont {E.}~\bibnamefont
  {Calzetta}}, \bibinfo {author} {\bibfnamefont {B.}~\bibnamefont {Hu}}, \ and\
  \bibinfo {author} {\bibfnamefont {E.}~\bibnamefont {Verdaguer}},\ }\href@noop
  {} {\bibfield  {journal} {\bibinfo  {journal} {International Journal of
  Modern Physics B}\ }\textbf {\bibinfo {volume} {21}},\ \bibinfo {pages}
  {4239} (\bibinfo {year} {2007})}\BibitemShut {NoStop}%
\bibitem [{\citenamefont {Zaremba}\ \emph {et~al.}(1999)\citenamefont
  {Zaremba}, \citenamefont {Nikuni},\ and\ \citenamefont
  {Griffin}}]{Zaremba_1999}%
  \BibitemOpen
  \bibfield  {author} {\bibinfo {author} {\bibfnamefont {E.}~\bibnamefont
  {Zaremba}}, \bibinfo {author} {\bibfnamefont {T.}~\bibnamefont {Nikuni}}, \
  and\ \bibinfo {author} {\bibfnamefont {A.}~\bibnamefont {Griffin}},\
  }\href@noop {} {\bibfield  {journal} {\bibinfo  {journal} {Journal of Low
  Temperature Physics}\ }\textbf {\bibinfo {volume} {116}},\ \bibinfo {pages}
  {277} (\bibinfo {year} {1999})}\BibitemShut {NoStop}%
\bibitem [{\citenamefont {Schmid}(1966)}]{Schmid_1966}%
  \BibitemOpen
  \bibfield  {author} {\bibinfo {author} {\bibfnamefont {A.}~\bibnamefont
  {Schmid}},\ }\href@noop {} {\bibfield  {journal} {\bibinfo  {journal} {Physik
  der Kondensierten Materie}\ }\textbf {\bibinfo {volume} {5}},\ \bibinfo
  {pages} {302} (\bibinfo {year} {1966})}\BibitemShut {NoStop}%
\bibitem [{\citenamefont {Berloff}\ \emph {et~al.}(2014)\citenamefont
  {Berloff}, \citenamefont {Brachet},\ and\ \citenamefont
  {Proukakis}}]{Berloff_2014}%
  \BibitemOpen
  \bibfield  {author} {\bibinfo {author} {\bibfnamefont {N.~G.}\ \bibnamefont
  {Berloff}}, \bibinfo {author} {\bibfnamefont {M.}~\bibnamefont {Brachet}}, \
  and\ \bibinfo {author} {\bibfnamefont {N.~P.}\ \bibnamefont {Proukakis}},\
  }\href@noop {} {\bibfield  {journal} {\bibinfo  {journal} {Proceedings of the
  National Academy of Sciences}\ }\textbf {\bibinfo {volume} {111}},\ \bibinfo
  {pages} {4675} (\bibinfo {year} {2014})}\BibitemShut {NoStop}%
\bibitem [{\citenamefont {Kida}\ and\ \citenamefont
  {Orszag}(1990)}]{KidaOrszag1990}%
  \BibitemOpen
  \bibfield  {author} {\bibinfo {author} {\bibfnamefont {S.}~\bibnamefont
  {Kida}}\ and\ \bibinfo {author} {\bibfnamefont {S.~A.}\ \bibnamefont
  {Orszag}},\ }\href {\doibase 10.1007/BF01065580} {\bibfield  {journal}
  {\bibinfo  {journal} {Journal of Scientific Computing}\ }\textbf {\bibinfo
  {volume} {5}},\ \bibinfo {pages} {85} (\bibinfo {year} {1990})}\BibitemShut
  {NoStop}%
\bibitem [{\citenamefont {Shukla}\ \emph {et~al.}(2019)\citenamefont {Shukla},
  \citenamefont {Mininni}, \citenamefont {Krstulovic}, \citenamefont
  {di~Leoni},\ and\ \citenamefont {Brachet}}]{Shukla2019}%
  \BibitemOpen
  \bibfield  {author} {\bibinfo {author} {\bibfnamefont {V.}~\bibnamefont
  {Shukla}}, \bibinfo {author} {\bibfnamefont {P.~D.}\ \bibnamefont {Mininni}},
  \bibinfo {author} {\bibfnamefont {G.}~\bibnamefont {Krstulovic}}, \bibinfo
  {author} {\bibfnamefont {P.~C.}\ \bibnamefont {di~Leoni}}, \ and\ \bibinfo
  {author} {\bibfnamefont {M.~E.}\ \bibnamefont {Brachet}},\ }\href {\doibase
  10.1103/physreva.99.043605} {\bibfield  {journal} {\bibinfo  {journal}
  {Physical Review A}\ }\textbf {\bibinfo {volume} {99}} (\bibinfo {year}
  {2019}),\ 10.1103/physreva.99.043605}\BibitemShut {NoStop}%
\bibitem [{\citenamefont {Estrada}\ \emph
  {et~al.}(2022{\natexlab{b}})\citenamefont {Estrada}, \citenamefont
  {Brachet},\ and\ \citenamefont {Mininni}}]{AmetteEstrada2022}%
  \BibitemOpen
  \bibfield  {author} {\bibinfo {author} {\bibfnamefont {J.~A.}\ \bibnamefont
  {Estrada}}, \bibinfo {author} {\bibfnamefont {M.~E.}\ \bibnamefont
  {Brachet}}, \ and\ \bibinfo {author} {\bibfnamefont {P.~D.}\ \bibnamefont
  {Mininni}},\ }\href {\doibase 10.1103/physreva.105.063321} {\bibfield
  {journal} {\bibinfo  {journal} {Physical Review A}\ }\textbf {\bibinfo
  {volume} {105}} (\bibinfo {year} {2022}{\natexlab{b}}),\
  10.1103/physreva.105.063321}\BibitemShut {NoStop}%
\bibitem [{\citenamefont {Mininni}\ \emph {et~al.}(2011)\citenamefont
  {Mininni}, \citenamefont {Rosenberg}, \citenamefont {Reddy},\ and\
  \citenamefont {Pouquet}}]{Mininni2011}%
  \BibitemOpen
  \bibfield  {author} {\bibinfo {author} {\bibfnamefont {P.~D.}\ \bibnamefont
  {Mininni}}, \bibinfo {author} {\bibfnamefont {D.}~\bibnamefont {Rosenberg}},
  \bibinfo {author} {\bibfnamefont {R.}~\bibnamefont {Reddy}}, \ and\ \bibinfo
  {author} {\bibfnamefont {A.}~\bibnamefont {Pouquet}},\ }\href {\doibase
  10.1016/j.parco.2011.05.004} {\bibfield  {journal} {\bibinfo  {journal}
  {Parallel Computing}\ }\textbf {\bibinfo {volume} {37}},\ \bibinfo {pages}
  {316} (\bibinfo {year} {2011})}\BibitemShut {NoStop}%
\bibitem [{\citenamefont {Fontana}\ \emph {et~al.}(2020)\citenamefont
  {Fontana}, \citenamefont {Bruno}, \citenamefont {Mininni},\ and\
  \citenamefont {Dmitruk}}]{Fontana_2020}%
  \BibitemOpen
  \bibfield  {author} {\bibinfo {author} {\bibfnamefont {M.}~\bibnamefont
  {Fontana}}, \bibinfo {author} {\bibfnamefont {O.~P.}\ \bibnamefont {Bruno}},
  \bibinfo {author} {\bibfnamefont {P.~D.}\ \bibnamefont {Mininni}}, \ and\
  \bibinfo {author} {\bibfnamefont {P.}~\bibnamefont {Dmitruk}},\ }\href@noop
  {} {\bibfield  {journal} {\bibinfo  {journal} {Computer Physics
  Communications}\ }\textbf {\bibinfo {volume} {256}},\ \bibinfo {pages}
  {107482} (\bibinfo {year} {2020})}\BibitemShut {NoStop}%
\bibitem [{\citenamefont {White}\ \emph {et~al.}(2014)\citenamefont {White},
  \citenamefont {Anderson},\ and\ \citenamefont {Bagnato}}]{White2014}%
  \BibitemOpen
  \bibfield  {author} {\bibinfo {author} {\bibfnamefont {A.~C.}\ \bibnamefont
  {White}}, \bibinfo {author} {\bibfnamefont {B.~P.}\ \bibnamefont {Anderson}},
  \ and\ \bibinfo {author} {\bibfnamefont {V.~S.}\ \bibnamefont {Bagnato}},\
  }\href {\doibase 10.1073/pnas.1312737110} {\bibfield  {journal} {\bibinfo
  {journal} {Proceedings of the National Academy of Sciences}\ }\textbf
  {\bibinfo {volume} {111}},\ \bibinfo {pages} {4719} (\bibinfo {year}
  {2014})}\BibitemShut {NoStop}%
\bibitem [{\citenamefont {Henn}\ \emph {et~al.}(2008)\citenamefont {Henn},
  \citenamefont {Seman}, \citenamefont {Seco}, \citenamefont {Olimpio},
  \citenamefont {Castilho}, \citenamefont {Roati}, \citenamefont
  {Magalh{\~{a}}es}, \citenamefont {Magalh{\~{a}}es},\ and\ \citenamefont
  {Bagnato}}]{Henn2008}%
  \BibitemOpen
  \bibfield  {author} {\bibinfo {author} {\bibfnamefont {E.~A.~L.}\
  \bibnamefont {Henn}}, \bibinfo {author} {\bibfnamefont {J.~A.}\ \bibnamefont
  {Seman}}, \bibinfo {author} {\bibfnamefont {G.~B.}\ \bibnamefont {Seco}},
  \bibinfo {author} {\bibfnamefont {E.~P.}\ \bibnamefont {Olimpio}}, \bibinfo
  {author} {\bibfnamefont {P.}~\bibnamefont {Castilho}}, \bibinfo {author}
  {\bibfnamefont {G.}~\bibnamefont {Roati}}, \bibinfo {author} {\bibfnamefont
  {D.~V.}\ \bibnamefont {Magalh{\~{a}}es}}, \bibinfo {author} {\bibfnamefont
  {K.~M.~F.}\ \bibnamefont {Magalh{\~{a}}es}}, \ and\ \bibinfo {author}
  {\bibfnamefont {V.~S.}\ \bibnamefont {Bagnato}},\ }\href {\doibase
  10.1590/s0103-97332008000200012} {\bibfield  {journal} {\bibinfo  {journal}
  {Brazilian Journal of Physics}\ }\textbf {\bibinfo {volume} {38}},\ \bibinfo
  {pages} {279} (\bibinfo {year} {2008})}\BibitemShut {NoStop}%
\bibitem [{\citenamefont {Smith}\ \emph {et~al.}(2011)\citenamefont {Smith},
  \citenamefont {Campbell}, \citenamefont {Tammuz},\ and\ \citenamefont
  {Hadzibabic}}]{Smith2011}%
  \BibitemOpen
  \bibfield  {author} {\bibinfo {author} {\bibfnamefont {R.~P.}\ \bibnamefont
  {Smith}}, \bibinfo {author} {\bibfnamefont {R.~L.~D.}\ \bibnamefont
  {Campbell}}, \bibinfo {author} {\bibfnamefont {N.}~\bibnamefont {Tammuz}}, \
  and\ \bibinfo {author} {\bibfnamefont {Z.}~\bibnamefont {Hadzibabic}},\
  }\href {\doibase 10.1103/physrevlett.106.250403} {\bibfield  {journal}
  {\bibinfo  {journal} {Physical Review Letters}\ }\textbf {\bibinfo {volume}
  {106}} (\bibinfo {year} {2011}),\ 10.1103/physrevlett.106.250403}\BibitemShut
  {NoStop}%
\bibitem [{\citenamefont {Giorgini}\ \emph {et~al.}(1996)\citenamefont
  {Giorgini}, \citenamefont {Pitaevskii},\ and\ \citenamefont
  {Stringari}}]{Giorgini1996}%
  \BibitemOpen
  \bibfield  {author} {\bibinfo {author} {\bibfnamefont {S.}~\bibnamefont
  {Giorgini}}, \bibinfo {author} {\bibfnamefont {L.~P.}\ \bibnamefont
  {Pitaevskii}}, \ and\ \bibinfo {author} {\bibfnamefont {S.}~\bibnamefont
  {Stringari}},\ }\href {\doibase 10.1103/physreva.54.r4633} {\bibfield
  {journal} {\bibinfo  {journal} {Physical Review A}\ }\textbf {\bibinfo
  {volume} {54}},\ \bibinfo {pages} {R4633} (\bibinfo {year}
  {1996})}\BibitemShut {NoStop}%
\bibitem [{\citenamefont {Davis}\ \emph {et~al.}(1995)\citenamefont {Davis},
  \citenamefont {Mewes}, \citenamefont {Andrews}, \citenamefont {van Druten},
  \citenamefont {Durfee}, \citenamefont {Kurn},\ and\ \citenamefont
  {Ketterle}}]{Davis1995}%
  \BibitemOpen
  \bibfield  {author} {\bibinfo {author} {\bibfnamefont {K.~B.}\ \bibnamefont
  {Davis}}, \bibinfo {author} {\bibfnamefont {M.~O.}\ \bibnamefont {Mewes}},
  \bibinfo {author} {\bibfnamefont {M.~R.}\ \bibnamefont {Andrews}}, \bibinfo
  {author} {\bibfnamefont {N.~J.}\ \bibnamefont {van Druten}}, \bibinfo
  {author} {\bibfnamefont {D.~S.}\ \bibnamefont {Durfee}}, \bibinfo {author}
  {\bibfnamefont {D.~M.}\ \bibnamefont {Kurn}}, \ and\ \bibinfo {author}
  {\bibfnamefont {W.}~\bibnamefont {Ketterle}},\ }\href {\doibase
  10.1103/physrevlett.75.3969} {\bibfield  {journal} {\bibinfo  {journal}
  {Physical Review Letters}\ }\textbf {\bibinfo {volume} {75}},\ \bibinfo
  {pages} {3969} (\bibinfo {year} {1995})}\BibitemShut {NoStop}%
\end{thebibliography}%

\appendix*
\section{Estimation of $T_\lambda$ \label{appendix}}

In order to estimate the critical temperature $T_\lambda$ of the condensate, we performed a series of simulations of the stochastic Ginzburg-Landau equation with parameters as listed in Sec.~\ref{sec:numerics} (i.e the most interacting case in our paper), at constant frequency of the trap $\omega_c$, varying the temperature $T$. In mean field theory, critical temperature decreases with increasing interaction, a result that has been confirmed experimentally where higher order corrections where also observed \cite{Smith2011, Giorgini1996}. This guarantees that temperatures used in this study, as they  are far below $T_\lambda$ of the most interacting case, are sufficiently small so that all the simulations are below critical temperature. This is enhanced by the fact that to determine $T_\lambda$ we use the smallest value of $\omega$ for the trap. To obtain the condensed fraction in the gas as a function of the temperature we use an approach similar to that used in experiments \cite{Davis1995}. We consider the mean mass density in the vicinity of the center of the trap as the order parameter, $\left< \rho_0 \right>$, averaged in time once the system reaches the equilibrium.

\begin{figure}[h]
    \centering
    \includegraphics[width=\columnwidth]{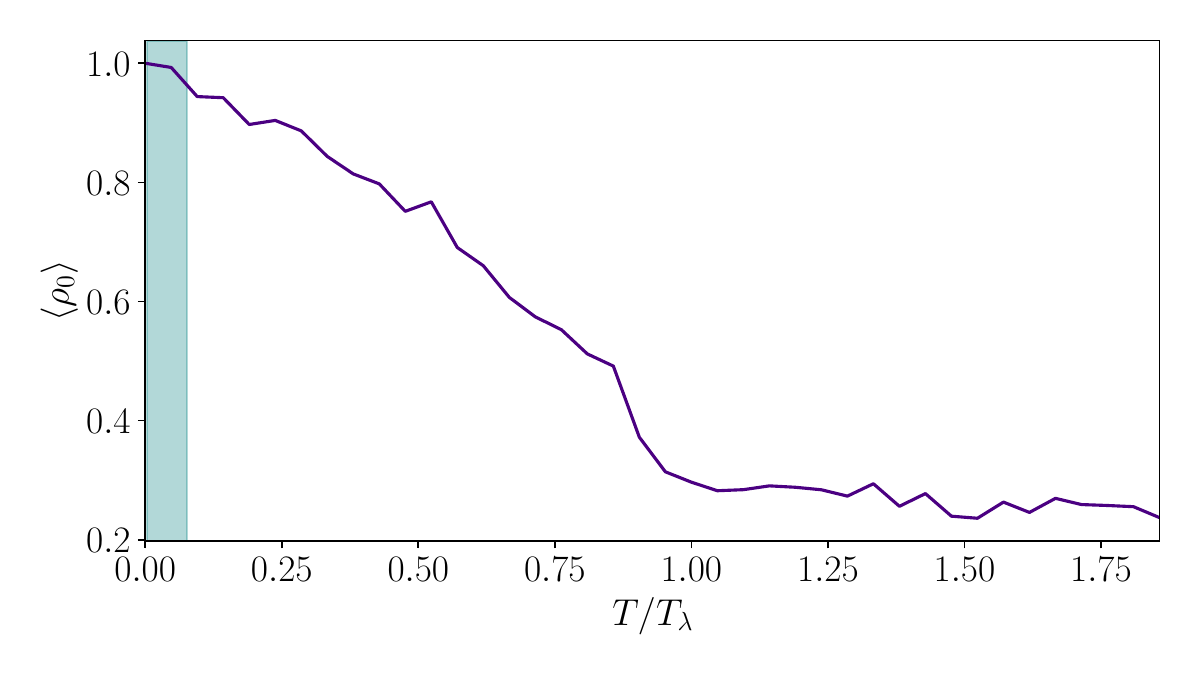}
    \caption{Mean value for different times of the mean density in the center of the trap as a function of the temperature for thermal states. The shaded part shows where all the temperatures mentioned in this work lay showing that they are far away from the transition so the non-condense fraction can be neglected.}
    \label{fig:critical temperature}
\end{figure}

Figure \ref{fig:critical temperature} shows the result as a function of the temperature. Temperature is shown in units of the transition temperatures $T_\lambda$, i.e., it is rescaled in such a way that the transition  in the behavior of $\left< \rho_0 \right>$ happens at $T/T_\lambda = 1$. Note that this parameter has an abrupt change of behavior at this temperature, as it decreases from 1 monotonically, and then remains approximately constant. This is compatible with a second order phase transition. All the temperatures considered in this study lie in the light-blue shaded region of Fig.~\ref{fig:critical temperature}, i.e., far away from the phase transition. 

\end{document}